\documentclass[12pt]{article}
\usepackage{graphicx}
\usepackage{epstopdf, epsfig}
\usepackage{amsfonts,amsmath,amssymb}
\usepackage[pdftex,colorlinks,citecolor=blue]{hyperref}
\usepackage{bm}
\usepackage{xcolor}
\usepackage{setspace}
\usepackage[figurename=Fig.]{caption}
\usepackage{lineno}

\usepackage[super,comma,sort&compress]{natbib}
\makeatletter
\newcommand*{\citenumns}[2][]{%
  \begingroup
  \let\NAT@mbox=\mbox
  \let\@cite\NAT@citenum
  \let\NAT@space\NAT@spacechar
  \let\NAT@super@kern\relax
  \renewcommand\NAT@open{}%
  \renewcommand\NAT@close{}%
  \cite[#1]{#2}%
  \endgroup
}
\makeatother

\usepackage{times}

\newenvironment{sciabstract}{%
\begin{quote} \bf}
{\end{quote}}

\newcommand{\affsize}{\footnotesize}
\newcommand{\Fu}{\bm{F}_{\perp}^{U}}
\newcommand{\Fb}{\bm{F}_{\perp}^{B}}
\newcommand{\ekwave}{\epsilon_{K,{\rm wave}}}

\newcommand{\pD}[2]{\frac{\partial #2}{\partial #1}}

\newcommand\bb[1]{\mbox{\boldmath{$#1$}}}
\newcommand\grad{\bb{\nabla}}
\newcommand\bcdot{\,\bb{\cdot}\,}
\newcommand\btimes{\,\bb{\times}\,}

\newcommand{\rmd}{{\rm d}}

\newcommand{\ez}{\hat{\bb{z}}}
\newcommand{\eb}{\hat{\bb{b}}}

\newcommand{\nkaw}{n_{\rm KAW}}

\usepackage[normalem]{ulem}

\newcommand{\revchng}[1]{{#1}}

\title{High-frequency heating of the solar wind triggered by low-frequency turbulence}

\author
{Jonathan~Squire,$^{1\ast}$ 
Romain Meyrand,$^{1}$ 
Matthew~W.~Kunz,$^{2,3}$\\
Lev Arzamasskiy,$^{4}$
 Alexander~A.~Schekochihin,$^{5,6}$
Eliot~Quataert$^{2}$ \\
\\
\affsize{$^{1}$Physics Department, University of Otago, Dunedin 9010, New Zealand,}\\
\affsize{$^{2}$Department of Astrophysical Sciences, Princeton University, Peyton Hall, Princeton, NJ 08544, USA}\\
\affsize{$^{3}$ Princeton Plasma Physics Laboratory, PO Box 451, Princeton, NJ 08543, USA}\\
\affsize{$^{4}$ School of Natural Sciences, Institute for Advanced Study, Princeton, NJ 08544, USA}\\
\affsize{$^{5}$ The Rudolf Peierls Centre for Theoretical Physics, University of Oxford, Clarendon Laboratory, Parks Road,}\\\affsize{ Oxford, OX1 3PU, UK}\\
\affsize{$^{6}$ Merton College, Oxford OX1 4JD, UK}\\
\scriptsize{$^\ast$To whom correspondence should be addressed; E-mail:  jonathan.squire@otago.ac.nz.}
}

\date{}

\begin{document} 


\baselineskip16pt


\maketitle 


\vspace{-0.5cm}
\begin{sciabstract}
The fast solar wind's high speeds and nonthermal features require that significant  heating occurs well above the Sun's surface.
Two leading theories seem incompatible: 
low-frequency ``Alfv\'enic'' turbulence, which transports energy outwards and  is observed ubiquitously by spacecraft but struggles to explain the observed dominance of ion over electron heating; and high-frequency ion-cyclotron waves (ICWs), which
 explain the   nonthermal heating of ions but lack an obvious source. Here, we argue that the recently proposed  ``helicity barrier'' effect, 
 which limits electron heating by inhibiting the turbulent cascade of energy to the smallest scales, can unify these two paradigms. Our six-dimensional  simulations show how the helicity barrier causes the 
large-scale energy to grow in time, generating small parallel scales and high-frequency ICW heating from  low-frequency turbulence, while simultaneously explaining various other long-standing observational puzzles.
The predicted causal link between plasma expansion and the ion-to-electron heating ratio suggests that the helicity 
barrier could contribute to key observed differences between fast- and slow-wind streams.
\end{sciabstract}


\vspace{0.5cm}
The basic mechanisms that heat the solar corona and accelerate the solar wind remain mysterious despite intensive study over
many decades\cite{Cranmer2019}. A successful theory must explain how energy contained in photospheric  motions and magnetic fields 
can be liberated to cause extreme and sudden heating
of the coronal plasma, along with its  acceleration to velocities well in excess of the escape velocity of the Sun.
Adding to the complexity, the coronal plasma is collisionless\,---\,the mean-free path of protons can be large compared 
to the largest observed structures\,---\,meaning it can be far out of local thermal equilibrium. This freedom opens up  a wide 
array of channels for plasma heating\,---\,ions might be heated more than electrons (or vice versa), or particles might gain energy
preferentially in a particular direction with respect to the local magnetic field \cite{Marsch2006}. Such differences can have significant macroscopic 
consequences. 

The dominant heating mechanism(s) must be consistent with an extensive array of measurements taken both remotely, from the low corona itself, and \emph{in situ},
from spacecraft spread throughout the solar wind. In fast-wind streams, these data indicate that the heating 
must be spatially extended out to several solar radii in order to drive observed wind speeds \cite{Parker1965}.
It must  preferentially heat protons over electrons \cite{Hansteen1995}, while heating heavier ions (e.g., alpha particles) even 
more effectively \cite{Kohl1997}. It must  heat protons  
in the direction perpendicular to the local magnetic field significantly more than in the parallel direction, in order to explain  temperature anisotropies \cite{Cranmer1999}. And,  its features and/or aftereffects should be observable 
in the measured field fluctuations and particle distribution, particularly 
at the low altitudes now being explored by Parker Solar Probe (PSP) \cite{Bale2019}. 

A  paradigm that can\,---\,at least in principle\,---\,satisfy the above requirements is heating through ``Alfv\'enic'' turbulence.
Low-frequency Alfv\'enic motions in the low corona are observed to contain sufficient energy to power the wind \cite{DePontieu2007,Tomczyk2007}, 
and there are well-developed theories for how such motions become turbulent following reflection from large-scale density gradients \cite{Velli1989,vanBallegooijen2011,Shoda2019}. This 
turbulence transfers  energy into successively smaller-scale 
motions perpendicular to the magnetic field (larger $k_{\perp}$, where $k_{\perp}$ is the inverse perpendicular scale), 
ultimately dissipating to  heat the plasma. The difficulty
is that most theories predict that, in the strongly magnetized limit relevant 
to the solar corona (the low-$\beta$ limit, where $\beta $ is the ratio of thermal to magnetic pressure), such low-frequency, high-$k_{\perp}$ structures dissipate to heat predominantly \emph{electrons} \cite{Quataert1999,Schekochihin2019}.
Other 
low-frequency plasma motions, such as compressive waves, generally cause parallel heating of ions \cite{Schekochihin2009}. Both 
possibilities are inconsistent with observations.  \revchng{More promisingly,} for turbulence of sufficient amplitudes, 
``stochastic heating''  \cite{Chandran2010} can  heat 
ions through a random walk on ion-gyroscale electric-field fluctuations. \revchng{Although it can plausibly 
explain key observations\cite{Chandran2011,Vech2017}, questions 
remain, such as its possible quenching due to  flattening of the distribution function\cite{Arzamasskiy2019,Cerri2021} and the influence of cross helicity\cite{Teaca2014}. }
\revchng{ Another possibility\,---\,that ions are heated by kinetic-Alfv\'en-wave (KAW) turbulence at sub-gyroradius scales \cite{Arzamasskiy2019,Isenberg2019}\,---\,remains  less well understood and may be inefficient at  low $\beta$ \cite{Cerri2021}}.
 
In the opposite limit of short field-parallel wavelengths (large $k_{\|}$), high-frequency ion-cyclotron waves (ICWs) 
provide a simpler mechanism to cause strong perpendicular ion heating \cite{Cranmer1999,Hollweg2002}.  At wavenumbers approaching $k_{\|}\sim d_{i}^{-1}$, where $d_{i}$ is the ion inertial 
length, their frequency approaches the ion gyrofrequency, where the cyclotron resonance causes 
highly efficient energy transfer from electromagnetic fields to ion velocities \cite{Kennel1966,Isenberg2011}. ICWs 
are observed ubiquitously \emph{in situ} \cite{Bale2019} and can suprathermally heat 
 minor ions in a way that is observationally compelling \cite{Kasper2013,Zhao2020}. 
 However, a sufficiently energetic direct solar source of ICWs
 is highly unlikely\cite{Hollweg2000}, and the Alfv\'enic cascade does not efficiently transfer energy to small parallel scales \cite{Shebalin1983}, seemingly ruling out their turbulent origin. Although their occurrence in data can be explained by kinetic instabilities\cite{Woodham2019},
in most theories this implies they would cool, rather than heat, the plasma\cite{Kennel1966}, except perhaps in the presence 
of strong nonthermal particle beams\cite{Voitenko2002}.

If combined, these two heating paradigms\,---\,via  Alfv\'enic turbulence or ICWs\,---\,can conceivably satisfy the fast-wind heating requirements described above, 
  maintaining an abundant source of perpendicular ion   heating well above the solar surface. 
 Here we assess whether a newly discovered effect, termed the ``helicity barrier''\cite{Meyrand2021}, can fulfil this role by  obstructing the dissipation of  collisionless Alfv\'enic  turbulence
 into electron heat. 
 Using  six-dimensional, high-resolution,  hybrid-kinetic simulations, we explore the 
effect of the helicity barrier on collisionless turbulent  heating, choosing parameters to match as closely as possible the conditions observed in fast-wind streams.
We assess the relevance of our results to the solar wind by comparing detailed features of the
turbulent spectra and ion distribution function to observations from PSP and other spacecraft. 

  \paragraph*{The helicity barrier}
  
Solar-wind
turbulence is \emph{imbalanced} (possessing cross helicity), meaning that it is energetically dominated by Alfv\'enic structures that  propagate outward from the Sun (designated $z^{+}$; the inward-propagating component is designated $z^{-}$).
The theory of highly perpendicular ($k_{\perp}\gg k_{\|}$) perturbations in a collisionless plasma
predicts that, at \revchng{scales larger than} the ion gyroradius $\rho_{i}$, imbalanced turbulent Alfv\'enic energy in $z^{\pm}$ can  cascade towards smaller scales (larger $k_{\perp}$),
as required to heat the plasma \cite{Schekochihin2009}. In constrast,
at sub-gyroradius scales $k_{\perp}\rho_{i}\gtrsim1$, magnetic helicity conservation implies the opposite\,---\,imbalanced energy must cascade \emph{inversely}, towards larger scales  \cite{Schekochihin2009,Cho2011,Meyrand2021}. The effect is directly analogous to 2D hydrodynamic turbulence, where enstrophy 
conservation causes energy to cascade inversely, creating large-scale vortex structures. 
However, unlike in hydrodynamics, in low-$\beta$ plasmas,  cross helicity at $k_{\perp}\rho_{i}\lesssim1$ transforms conservatively into magnetic helicity at $k_{\perp}\rho_{i}\gtrsim1$  
(the system conserves a generalized helicity \cite{Meyrand2021}). This implies that an imbalanced energy flux arriving at $k_{\perp}\rho_{i}\sim1$ from large scales cannot cascade to arbitrarily  small scales.

Mathematically, it is helpful to separate the turbulent energy flux $\varepsilon$ into 
components associated with the large-scale outward ($\varepsilon^{+}$) and inward  ($\varepsilon^{-}$) propagating fluctuations: $\varepsilon=\varepsilon^{+}+\varepsilon^{-}$. 
Generalized-helicity conservation prohibits the conversion of  $\varepsilon^{+}$ into $\varepsilon^{-}$ at any scale (note, however, that at $k_{\perp}\rho_{i}\gtrsim 1$, $\varepsilon^+$ is associated with a mixture of  outward- and inward- propagating KAWs). Because a forward   
 cascade must be balanced  ($\varepsilon^{+}\approx\varepsilon^{-}$) at $k_{\perp}\rho_{i}\gtrsim 1$, 
 only a small portion ${\sim}2\varepsilon^{-}$ of the energy flux can cascade to small scales where it will heat electrons \cite{Schekochihin2019}. The rest of the flux, ${\sim} (\varepsilon^{+}-\varepsilon^{-})$,
 is stuck\,---\,it hits the ``helicity barrier'' and thus remains at scales $k_{\perp}\rho_{i}\lesssim1$.  If
the system is forced continuously, this large-scale energy grows in time with a decreasing
parallel correlation length \cite{Meyrand2021}, as expected from critical balance \cite{Schekochihin2009}. 
We  show that this growth  eventually funnels the turbulent
energy into a spectrum of ICW fluctuations, heating the ions, which absorb the majority of the energy flux. 

 \subsection*{Numerical method}

Our simulation uses the \textsc{Pegasus}\texttt{++} code \cite{Kunz2014a}, which solves the hybrid-kinetic equations with isothermal electrons using 
the particle-in-cell method. The system is strongly magnetized with mean magnetic field $\bm{B}_{0}=-B_{0}\hat{\bm{z}}$, Alfv\'en speed $v_{\rm A}=B_{0}/\sqrt{4\pi n m_{i}}$,   and initial ion $\beta$, $\beta_{i}\equiv8\pi n k_{B}T_{i}/B_{0}^{2}=0.3$ ($m_{i}$, $n$, and $k_{B}T_{i}=m_{i}v_{\rm th}^{2}/2$  are the ion mass, number density, and  temperature, respectively with $v_{\rm th}$ the ion thermal velocity and $k_{B}$  the Boltzmann constant). Perpendicular ($x$ and $y$ directed) ion-velocity fluctuations
$\bm{u}_{\perp}$ and magnetic fluctuations $\bm{B}_{\perp}$ are driven at large scales and correlated to create imbalance, with $\bm{z}^{+}\equiv \bm{u}_{\perp}+ \bm{B}_{\perp}/\sqrt{4\pi nm_{i}}\gg \bm{z}^{-}\equiv \bm{u}_{\perp}- \bm{B}_{\perp}/\sqrt{4\pi nm_{i}}$, \revchng{($\bm{z}^{+}$  perturbations propagate in the $+\hat{\bm{z}}$ direction)}. The energy-injection rate $\varepsilon$ and cross-helicity 
injection rate $\varepsilon_{H} = \varepsilon^{+}- \varepsilon^{-}=0.9\varepsilon$ are constant in time.
Plasma heating is strongly influenced by the  amplitude $\delta B_{\perp}/B_{0}$  and spectral anisotropy 
$k_{\|}/k_{\perp}$ of fluctuations with $k_{\perp}\rho_{i}\sim 1$.
In order to reach realistic values  without simulating the enormous scale separation of the real 
solar wind, we use a highly elongated box with dimensions $L_{\|}=L_{z}=6L_{\perp}$; given the box size $k_{\perp0}\rho_{i}\equiv 2\pi\rho_{i}/L_{\perp}\approx0.05$, 
this gives conditions near $k_{\perp}\rho_{i}\sim 1$ that are comparable to those observed  \cite{Chen2016a}.
The elongated domain also implies that the timescales we probe are rapid compared to 
the solar wind's outer scales and its expansion rate, justifying  the external forcing to represent a turbulent flux from larger 
scales\cite{Davidson2004} and our neglect of expansion effects.
The simulation's resolution is $N_{\perp}^{2}\times N_{\|}=392^{2}\times 2352$ cells, so that the smallest resolved scales have $k_{\perp,{\rm max}}\rho_{i}\simeq \pi N_{\perp}\rho_{i}/L_{\perp} = 10 $. 
Most other simulation parameters, including a (hyper-)resistivity that dissipates small-scale magnetic energy, are chosen 
 to match a  previous  balanced turbulence simulation with $\beta_{i}=0.3$ \cite{Arzamasskiy2019}. This allows direct comparison of their spectra and heating.
Further  details are provided in Methods.

 \subsection*{Results}

\begin{figure}[!htb]
\centering
\includegraphics[width=0.55\columnwidth]{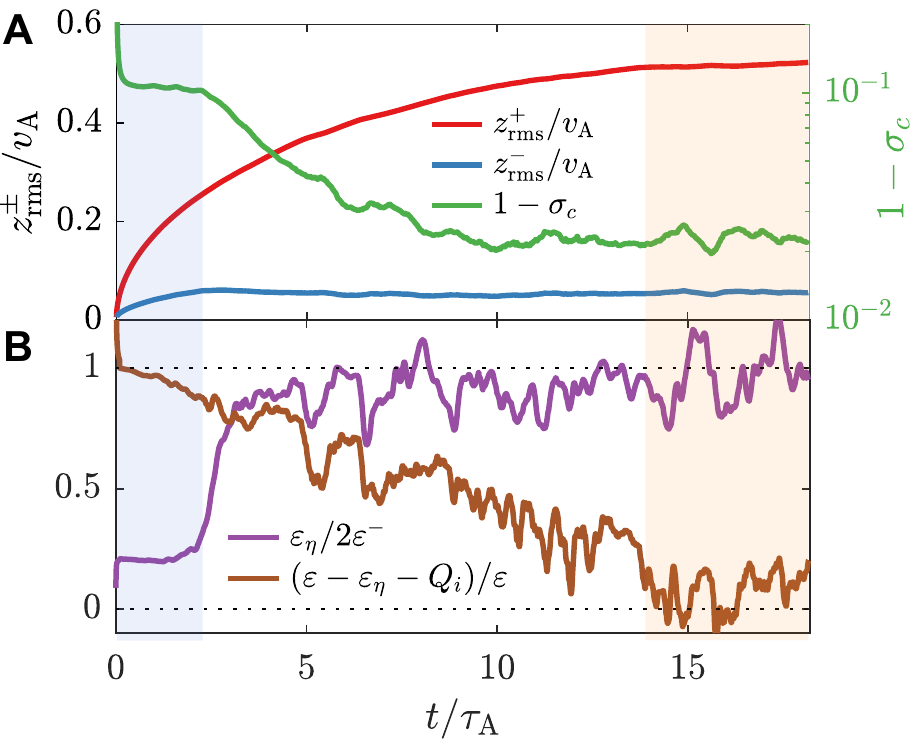}
\caption{\textbf{The time evolution confirms the formation of the helicity barrier}. (\textbf{A}) \revchng{Outward/inward-propagating} fluctuation amplitudes ($z^{\pm}_{\rm rms}$) and imbalance ($\sigma_{c}$); (\textbf{B}) Energy budget and heating, \revchng{illustrated through the  ion heating rate ($Q_{i}$) and the small-scale resistive dissipation ($\varepsilon_{\eta}$)}. The balanced
part of the injected energy (${\sim}2\varepsilon^{-}$) saturates early by $t\simeq3\tau_{\rm A}=3L_{\|}/v_{\rm A}$ (blue shaded region); \revchng{this is demonstrated by the saturation of  $z^{-}$ (blue line in \textbf{A}) and the  resistive dissipation (purple line in \textbf{B}), which absorbs only a small fraction
of the input energy $\varepsilon_{\eta}\approx 2\varepsilon^{-}\ll\varepsilon$.} Eventually, by $t\approx 14\tau_{\rm A}$ (orange shaded region), \revchng{ion heating absorbs the remainder of the injected energy so that $Q_{i}\approx \varepsilon-\varepsilon_{\eta}$ (brown line in \textbf{B}), halting the growth 
of $z^{+}$.} A numerical cooling effect has been removed  to compute the energy budget (see Methods and Extended Data Fig.~1).}
\label{fig: 1}
\end{figure}

The simulation's time evolution, shown in Fig.~1, 
exhibits several features that are expected from the helicity barrier \cite{Meyrand2021} but not from other theories of imbalanced turbulence \cite{Schekochihin2020,Meyrand2021}. 
Panel A shows the growth of the root-mean-square ampliutudes $z^{\pm}_{\rm rms}/v_{\rm A}\equiv\langle (z^{\pm})^{2}\rangle^{1/2}/v_{\rm A}$ 
and  imbalance (normalized cross helicity) $\sigma_{c} =2\langle \sqrt{4\pi nm_{i}}\bm{u}\bm{\cdot}\bm{B}\rangle/\langle 4\pi nm_{i} \bm{u}^{2} + \bm{B}^{2} \rangle$ (where $\langle \,\dots\,\rangle$ denotes a box average). Because $z^{+}_{\rm rms}/v_{\rm A}\approx 2 u_{\perp,{\rm  rms}}/v_{\rm A}\approx 2 B_{\perp,{\rm  rms}}/B_{0} $, the final   $\delta B_{\perp}/B_{0}\approx 0.26$ is nearly twice that of the balanced simulation ($\delta B_{\perp}/B_{0}\approx 0.14$). We see that $z^{-}$ saturates  quickly, by time $t\simeq 3\tau_{\rm A}$, while saturation of $z^{+}$ occurs only after $t\simeq 14\tau_{\rm A}$ ($\tau_{\rm A}\equiv  L_{\|}/v_{\rm A}$ is the Alfv\'en time). This is expected: energy in $z^{-}$ can cascade to $k_{\perp}\rho_{i}\gtrsim1$ through standard  KAW turbulence, while most of the energy in $z^{+}$ cannot cascade past $k_{\perp}\rho_{i}\sim1$ due to the helicity barrier. 
The imbalance saturates at $\sigma_{c}\approx0.98$, which, though large, is regularly observed by PSP \cite{McManus2020}.
 Panel B shows the ion heating rate $Q_{i}=V\partial_{t}\langle 3m_{i} v_{\rm th}^{2}/2\rangle$ (where $V$ is the simulation volume)
and the (hyper-)resistive dissipation rate $\varepsilon_{\eta}$. The latter is a proxy for electron heating because 
it absorbs the energy  that cascades to the smallest scales below where ions can respond to the fluctuations. 
We see that $\varepsilon_{\eta}$ saturates rapidly, together with $z^{-}$ and well before the total energy. Its value, $\varepsilon_{\eta}\approx2\varepsilon^{-}$ approximates the balanced portion of the injected flux, as expected. The remainder of the energy input, $\varepsilon - \varepsilon_{\eta}$, must 
eventually go into ion heating $Q_{i}$. However, in order for this to happen, $z^{+}$ must grow significantly, which, notably,
occurs without change to $\varepsilon_{\eta}$ or $z^{-}$.
The simulation saturates with $Q_{i}\approx \varepsilon - \varepsilon_{\eta}$ at $t\approx 14\tau_{\rm A}$; we run it for another ${\approx}4.2\tau_{\rm A}$ in this steady state. 

\begin{figure}[!htb]
\centering
\includegraphics[width=0.75\columnwidth]{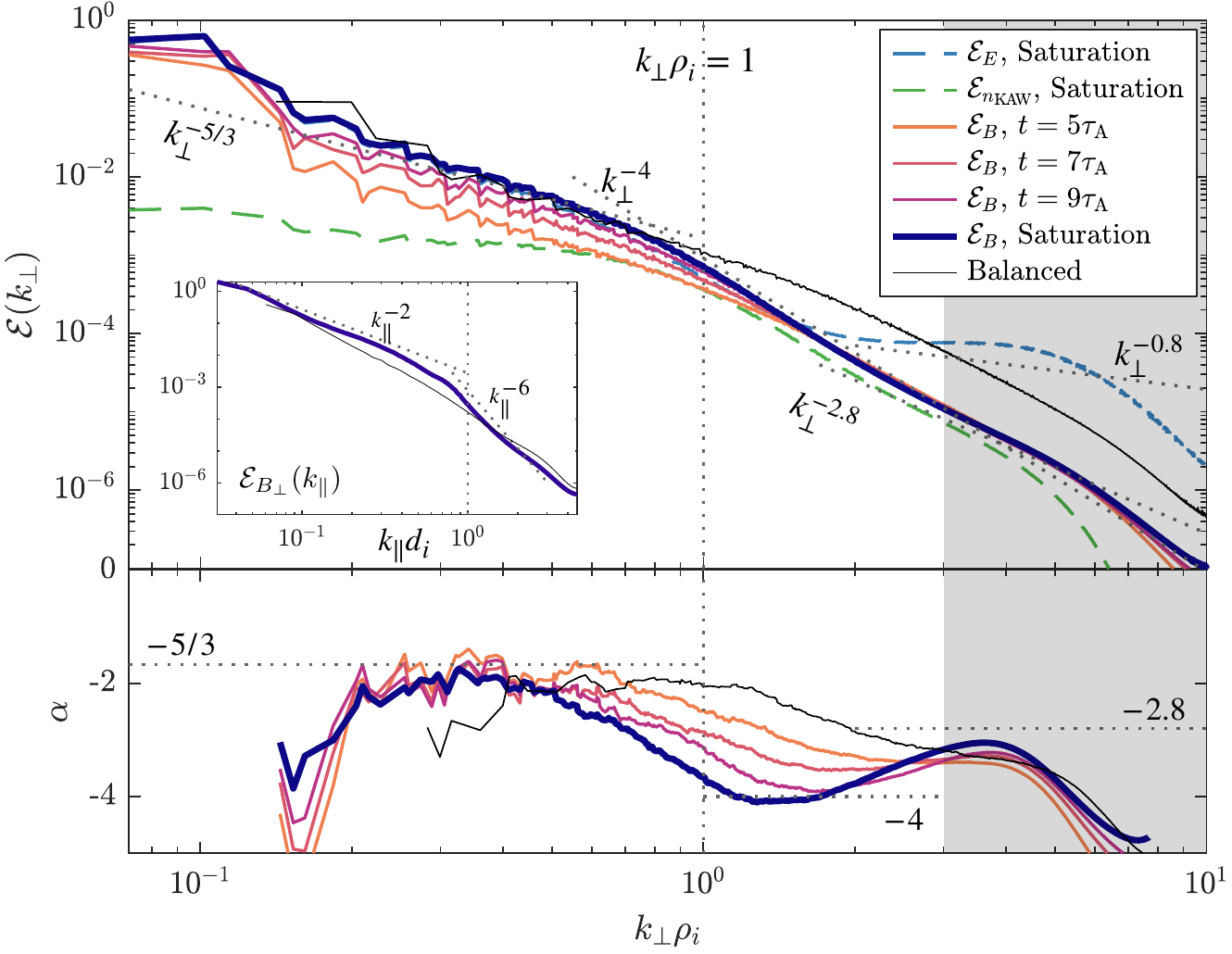}
\caption{\textbf{Fluctuation spectra exhibit a steep transition region around $k_{\perp}\rho_{i}=1$.} Spectra of the  magnetic field 
$\mathcal{E}_{\bm{B}}= \mathcal{E}_{B_{x}}+\mathcal{E}_{B_{y}}+\mathcal{E}_{B_{z}}$, electric field $\mathcal{E}_{\bm{E}}= \mathcal{E}_{E_{x}}+\mathcal{E}_{E_{y}}+\mathcal{E}_{E_{z}}$, and 
KAW-normalized density $\nkaw=\sqrt{\beta_{i}(1+2\beta_{i})}n$, binned in $k_{\perp}=\sqrt{k_{x}^{2}+k_{y}^{2} }$. 
The lower panel shows the local scaling exponents, $\alpha = {\rm d}\ln\mathcal{E}/{\rm d}\ln k_{\perp}$; at small scales (approximately the shaded region) spectra are affected by particle noise (see Extended Data Fig.~2). The inset shows  the extremely steep $k_{\|}$ spectrum of $B_{\perp}$  ($\mathcal{E}_{B_{\perp}}= \mathcal{E}_{B_{x}}+\mathcal{E}_{B_{y}}$) near $k_{\|}d_{i}=1$. Thin black lines show $\mathcal{E}_{\bm{B}}$ for an  analogous balanced-turbulence
simulation \cite{Arzamasskiy2019}. }
\end{figure}

Perpendicular magnetic ($\mathcal{E}_{\bm{B}}$), electric ($\mathcal{E}_{\bm{E}}$), and density ($\mathcal{E}_{\nkaw}$) fluctuation spectra are shown in Fig.~2. The signature of the helicity barrier is the sharp spectral break at $k_{\perp}\rho_{i}<1$ in $\mathcal{E}_{\bm{B}}$ and $\mathcal{E}_{\bm{E}}$, which migrates to larger scales with time \cite{Meyrand2021}.  ``Double-kinked'' spectra\,---\,a  ${\sim}k_{\perp}^{-5/3}$ range at large scales, 
 a steep ${\sim} k_{\perp}^{-4}$ transition range bracketing $k_{\perp}\rho_{i}=1$, 
and  a flatter range of KAW turbulence (${\sim}k_{\perp}^{-2.8}$) at yet smaller scales\,---\,have been observed for decades\cite{Leamon1998} including by PSP close to the Sun\cite{Bowen2020a}, but have lacked a clear theoretical explanation. 
So far as we are aware, this is the first self-consistent kinetic simulation to exhibit this feature. Although  $\mathcal{E}_{\bm{B}}$ in Fig.~2 lacks a ${\sim}k_{\perp}^{-2.8}$ range due to box resolution (resistivity), it clearly re-flattens at $k_{\perp}\rho_{i}\approx2$, 
and the width of the transition range is comparable to observations \cite{Bowen2020a}. 
The  $\mathcal{E}_{\bm{E}}$ spectrum has a similar shape,  flattening after a steep drop around $k_{\perp}\rho_{i}\sim1$. 
The spectrum of $\nkaw=\sqrt{\beta_{i}(1+2\beta_{i})}n$ is predicted \cite{Schekochihin2009} and observed \cite{Chen2013} to satisfy $\mathcal{E}_{\nkaw}\approx \mathcal{E}_{\bm{B}}$
in KAW turbulence because  linear KAWs satisfy $\delta B_{\perp}/B_{0}\approx (\delta n/n)\sqrt{\beta_{i}(1+\beta_{i})}\approx (\delta B_{\|}/B_{0})\sqrt{1+1/\beta_{i}}$ and $\delta B^{2}=\delta B_{\perp}^{2}+\delta B_{\|}^{2}$\,---\,this occurs here for $k_{\perp}\rho_{i}\gtrsim1$, providing further evidence for the KAW-like nature of the sub-$\rho_{i}$ turbulence. 
The parallel spectrum (see Fig.~2 inset and Methods) exhibits a ${\sim} k_{\|}^{-2}$ range
at large scales, followed by a very steep ${\sim}k_{\|}^{-6}$ range bracketing $k_{\|}d_{i}\sim 1$, which flattens at smaller scales; again
these features match recent PSP observations \cite{Duan2021}. Balanced turbulence\cite{Arzamasskiy2019}, unlike the imbalanced case,  does not
exhibit a steep transition range in either $k_{\perp}$ or $k_{\|}$, \revchng{which is likely also the case in the solar wind\cite{Huang2021}}.
Note that  spectra are adversely affected by particle noise for $k_{\perp}\rho_{i}\gtrsim3$ (see Methods and Extended Data figures 1-2).

\begin{figure}
\centering
\includegraphics[width=1.0\columnwidth]{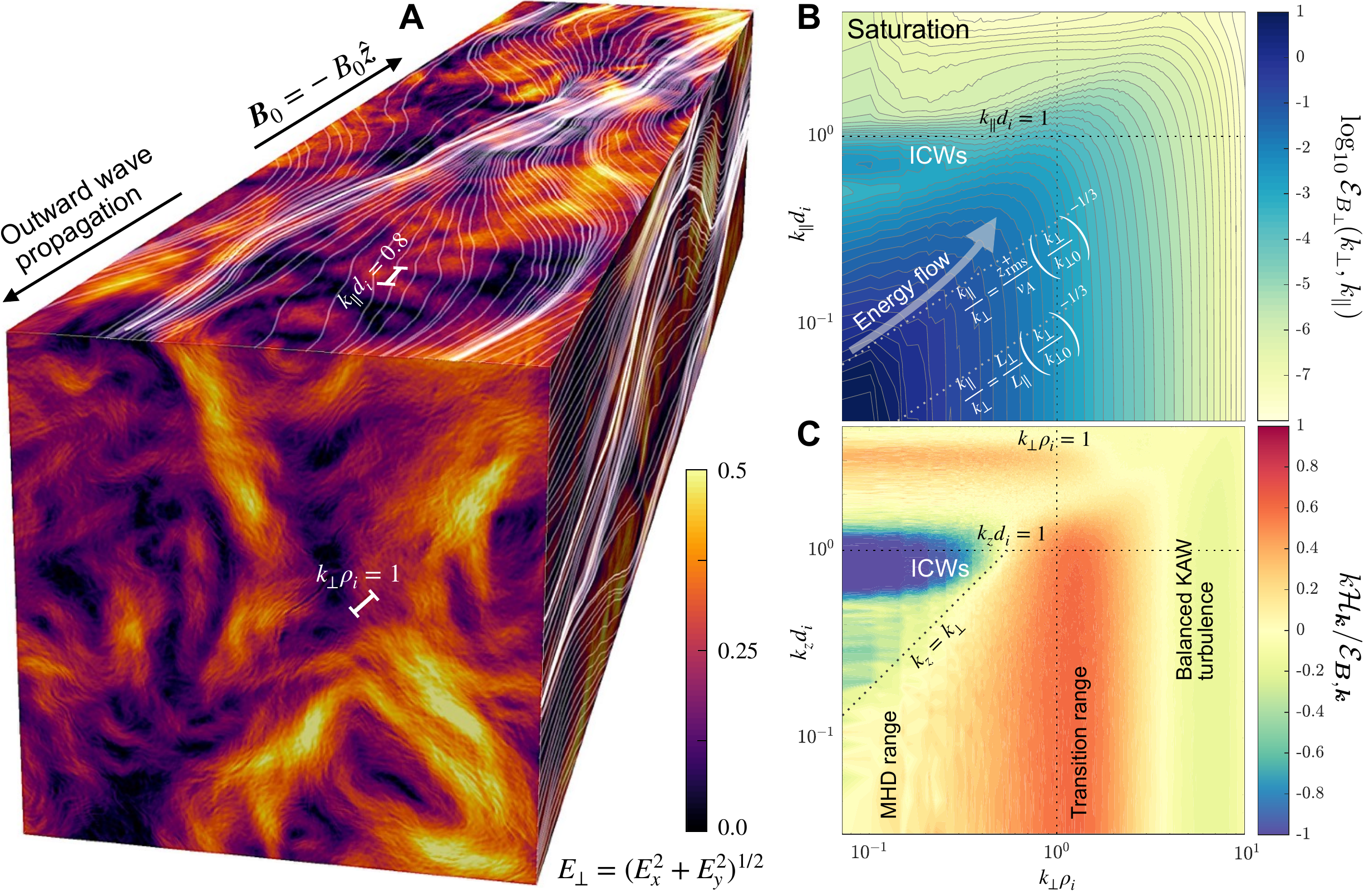}
\caption{\textbf{Evidence for ICW fluctuations in the saturated state.} (\textbf{A}) Structure of $E_{\perp}$;  superimposed 
white lines show magnetic-field lines
projected onto the plane. 
(\textbf{B}) Perpendicular  magnetic-fluctuation spectrum ($\mathcal{E}_{B_{\perp}}=\mathcal{E}_{B_{x}}+\mathcal{E}_{B_{y}}$) in ($k_{\perp},\,k_{\|}$), illustrating the path of energy to $k_{\|}d_{i}\sim 1$ scales (labelled ``Energy flow'') and the 
parallel ICW bump at $k_{\|}d_{i}\simeq 0.8$. The angled white dotted lines show the canonical  critical balance result\cite{Schekochihin2009} based on $z^{+}_{\rm rms}$ (top) or the box outer scale (bottom).  (\textbf{C}) Normalized magnetic helicity spectrum $k\mathcal{H}_{\bm{k}}/\mathcal{E}_{\bm{B}}$(see text)  in ($k_{\perp},\,k_{z}$), showing the reversed helicity signatures of KAWs ($k_{\perp}\gg k_{\|}$) and ICWs ($k_{\|}\gg k_{\perp}$). For perpendicular structures, $\mathcal{H}_{\bm{k}}\approx 0$ at both large scales  ($k_{\perp}\rho_{i}\ll1$), because large-scale Alfv\'en waves are nonhelical, and in the KAW range ($k_{\perp}\rho_{i}\gg1$), because the  KAW turbulence must be balanced. In between, in the transition range around $k_{\perp}\rho_{i}\simeq1$, $\mathcal{H}_{\bm{k}}$ has a local maximum because the Alfv\'enic fluctuations are both dispersive and imbalanced.   }
\label{fig: 3}
\end{figure}

Evidence for the presence of ICWs in the saturated state is shown in Fig.~3. 
The projected magnetic-field lines and  electric field (panel A) reveal the coexistence of $k_{\|}d_{i}\sim 1$  parallel structure with the sub-$\rho_{i}$ striations of  KAW turbulence in the perpendicular plane. 
More quantitatively, panel B shows the two-dimensional $(k_{\perp},k_{\|})$ spectrum of $B_{\perp}$ (see Methods and Extended Data figure 3).
Several features are manifest: first, unlike in the analogous balanced-turbulence simulation  \cite{Arzamasskiy2019}, the outer-scale parallel correlation length is significantly smaller than $L_{\|}$
because it decreases with increasing 
amplitude to maintain critical balance, $k_{\|}v_{\rm A} \sim k_{\perp} z^{+}$ (the region of maximal spectral power moves upwards in time
as $z^{+}$ grows);
second, the cone of maximal spectral power appears to steepen as the turbulence moves to smaller scales,  creating  small parallel scales faster than 
the canonical result\cite{Schekochihin2009} $k_{\|}\propto k_{\perp}^{2/3}$ \revchng{(or  $k_{\|}\propto k_{\perp}^{1/2}$ for aligned turbulence\cite{Schekochihin2020})};
third, there is a clear spectral bump at $k_{\|}d_{i}\simeq 0.8$ and $k_{\perp}<k_{\|}$, \revchng{the signature of ICWs. By  integrating the energy  spectrum over modes with $k_{\perp}\leq k_{\|}$, we
estimate that these ICW modes contain ${\simeq}1\%$ of the total energy; this fraction  grows by a factor ${\simeq}50$ from $t\simeq5\tau_{\rm A}$ to saturation and exceeds that of saturated balanced turbulence by a factor ${\simeq}30$.}
\revchng{We confirm that these modes are indeed ICWs in panel C}, which shows the normalized magnetic-helicity spectrum $k\mathcal{H}_{\bm{k}}/\mathcal{E}_{\bm{B}}$, where $\mathcal{H}_{\bm{k}} = i(B_{x}^{*}B_{y}-B_{y}^{*}B_{x})/k_{z}$, binned in ($k_{\perp},\,k_{z}$). As predicted \cite{Howes2010} and observed \cite{Huang2020}, 
 ICWs  (with $k_{\perp}\lesssim k_{\|}$) are characterised by $k\mathcal{H}_{\bm{k}}/\mathcal{E}_{\bm{B}}<0$,
while $k\mathcal{H}_{\bm{k}}/\mathcal{E}_{\bm{B}}>0$ for $k_{\perp}\gtrsim k_{\|}$ near $k_{\perp}\rho_{i}\simeq1$. This  results from  the intrinsic polarization  of 
ICWs and their propagation direction; $k\mathcal{H}_{\bm{k}}/\mathcal{E}_{\bm{B}}\approx -1$ indicates that  ICWs propagate almost exclusively in the $+\bm{\hat{z}}$ direction like the large-scale $z^{+}$.
At $k_{\perp}>k_{\|}$, $k\mathcal{H}_{\bm{k}}/\mathcal{E}_{\bm{B}}>0$ is a result of  the dominantly Alfv\'enic perturbations 
becoming dispersive at $k_{\perp}\rho_{i}\sim1$; helicity grows to $k\mathcal{H}_{\bm{k}}/\mathcal{E}_{\bm{B}}\approx 0.5$ at $k_{\perp}\rho_{i}\simeq1$
before decreasing again
because the small-scale KAW cascade is  balanced. This 
feature, which is a theoretical corollary of the helicity barrier, has been  commonly observed by PSP and other spacecraft \cite{Huang2020} \revchng{and correlates with the transition-range spectral slope as predicted\cite{Zhao2021}}.

Evidence that plasma heating occurs through ICWs is provided in  Fig.~4. Quasi-linear cyclotron-heating theory \cite{Kennel1966} is
based on the idea that ions and ICWs interact strongly if the wave frequency $\omega_{\bm{k}}$ is resonant with the Doppler-shifted
ion gyromotion. When ICWs exist across a range of $k_{\|}$, the process  flattens the  ion distribution function $f_{i}(w_{\perp},w_{\|})$
along specific ``scattering contours,'' which can be computed\cite{Isenberg2011} from $\omega_{\bm{k}}$ for waves
of a particular $k_{\perp}$  ($w_{\perp}$ and $w_{\|}$ are the  field-perpendicular and -parallel velocities of ions in the frame moving with the plasma).
Theory suggests that, because the scattering contours steepen with increasing $k_{\perp}/k_{\|}$, heating by oblique ICWs 
generates an $f_{i}$ that increases along the scattering contours of parallel ICWs \cite{Chandran2010a}. The consequences are twofold: first, 
oblique-ICW heating generates parallel ICWs, \revchng{explaining the dominance of $k_{\perp}\ll k_{\|}$ modes in Fig.~3B}; second, in quasi-steady state, $f_{i}$ is
nearly flat along the \emph{parallel} ICW scattering contours. We plot  these, along with $f_{i}(w_{\perp},w_{\|})$, in Fig.~4A. There is exceptionally good agreement at saturation for $w_{\|}\lesssim -w_{\|,d_{i}}$ particles,  which are those that can resonate with $k_{\|}d_{i}\lesssim 1$ 
ICWs propagating in the $+\hat{\bm{z}}$ direction (see Methods). The time evolution is also telling: 
the quasi-linear flattening starts at large $|w_{\|}|$ and moves upwards as time advances, which is expected because there is more power 
in low-$k_{\|}$ modes that resonate with high-$|w_{\|}|$ ions.
Panel B shows the perpendicular 
energy diffusion coefficient $D^{E}_{\perp\perp}$, computed for $w_{\|}<-w_{\|,d_{i}}$ using the time evolution of $f_{i}$\cite{Vasquez2020}  and validated
by  computing $\langle \bm{E}_{\perp}\bm{\cdot} \bm{w}_{\perp}\rangle$ directly from particle trajectories \cite{Cerri2021}. Quasi-linear ICW theory predicts\cite{Kennel1966} $D^{E}_{\perp\perp}\propto w_{\perp}^{2}$, as seen in Fig.~4, while a stochastically heated plasma has\cite{Cerri2021} $D^{E}_{\perp\perp}\sim {\rm const}$ for $w_{\perp}\sim v_{\rm th}$. This quantitatively confirms the dominance of quasi-linear ICW heating and we find no other evidence 
for stochastic heating in this simulation, \revchng{although it is possible that 
it could govern saturation under different conditions} (e.g., lower $\beta$; see Methods and Extended Data figure 4).
Finally, we see a clear  flattening of  $f_{i}(w_{\perp},w_{\|})$ at $w_{\|}\approx v_{\rm A}$ and small $w_{\perp}$, which forms a modestly super-Alfv\'enic  beam feature in the direction of dominant wave propagation,
similar to those observed in the fast solar wind  \cite{Marsch1982,Marsch2006,Verniero2020}.
By comparing $\partial f_{i}/\partial t$ with  $\langle {E}_{\|} {w}_{\|}\rangle$ (see Extended Data figure 5), we confirm that this arises through Landau damping of Alfv\'en waves as their phase velocity increases  near $k_{\perp}\rho_{i}\sim 1$ \cite{Li2010}. 

\begin{figure}
\centering
\includegraphics[width=1.0\columnwidth]{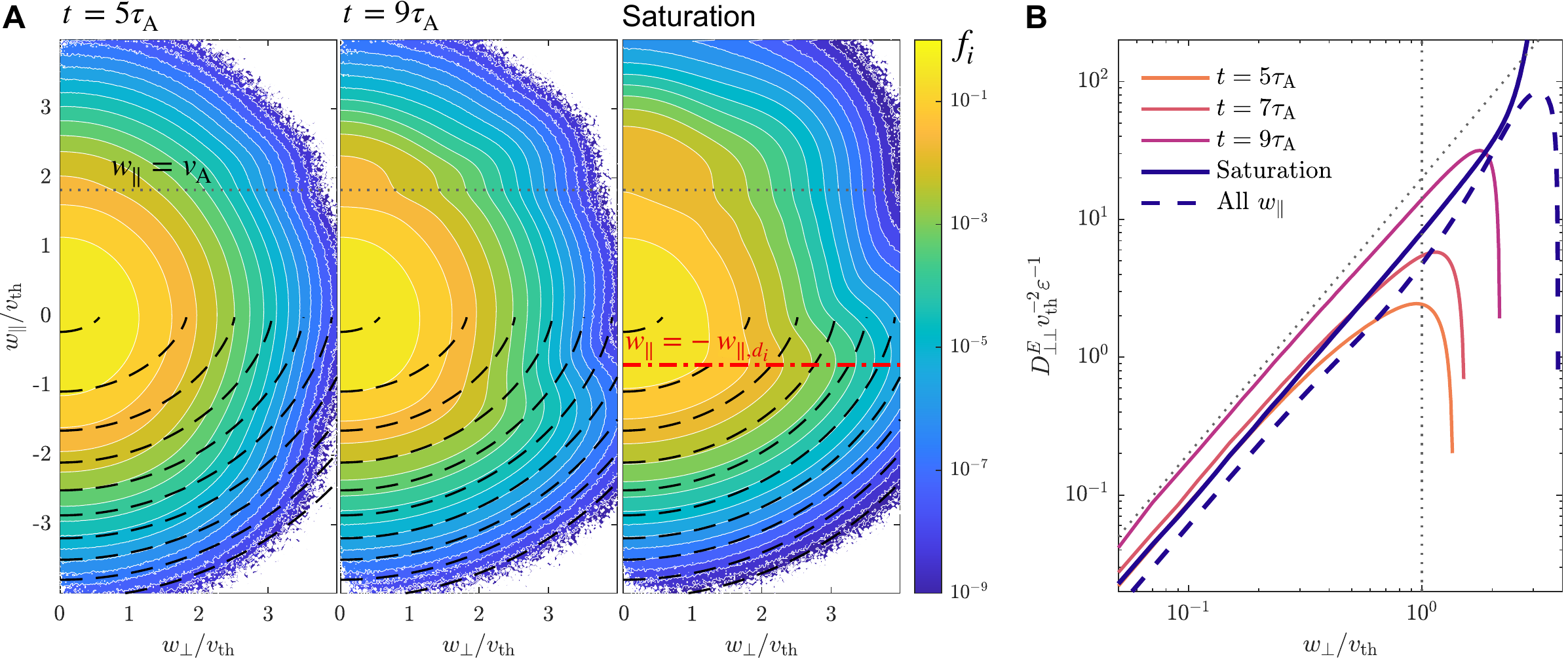}
\caption{\textbf{The evolution of the ion distribution function $f_{i}(w_{\perp},w_{\|})$ shows how ICWs heat  ions.} (\textbf{A}) Structure of $f_{i}$ at different times. 
At saturation,  $f_{i}$ aligns nearly perfectly with the scattering contours of parallel ICWs (dashed lines) for  $w_{\|}\lesssim -w_{\|,d_{i}}$ (red line), which is the region of velocity space where there exists resonant ICWs.  (\textbf{B}) Perpendicular energy diffusion coefficient $D^E_{\perp\perp}$; 
solid lines include only $w_{\|}< -w_{\|,d_{i}}$; the dashed line includes all $w_{\|}$, with the lower values of $D^E_{\perp\perp}$ indicating less heating for $w_{\|}>0$. }
\label{fig: 4}
\end{figure}

 \subsection*{Discussion}

Together, Figs.~3 and 4 provide strong evidence for heating through ICWs generated by
 imbalanced Alfv\'enic turbulence regulated by the helicity barrier. 
Given the excellent match to \emph{in-situ} observations of spectra, helicity, and distribution functions, we suggest the same is true in the fast solar wind, 
 reconciling the paradigms of low-frequency Alfv\'enic turbulence and ICW heating. If,
  as suggested above,  ICWs are produced due to the decrease in outer parallel scale with increasing amplitude, 
then the turbulence will saturate once it has grown sufficiently for $k_{\|}d_{i}\sim1$ scales to be reached (where Alfv\'en waves become ICWs)
before the  $k_{\perp}\rho_{i}\sim1 $ scales  (where Alfv\'en waves become KAWs). If large-scale fluctuations are critically balanced 
 with spectrum $\mathcal{E}_{\bm{B}}\propto k_{\perp}^{-5/3}$, this suggests a critical outer-scale (${\sim}L_{\perp}$) saturation amplitude for appreciable ICW heating given by  $\delta z^{+}/v_{\rm A}\sim A \beta_{i}^{1/2} (\rho_{i}/L_{\perp})^{-1/3}$, with  our simulation giving a proportionality factor $A\simeq0.2$ (an observationally testable prediction).
\revchng{A corollary is that the energy-injection rate sets only the timescale to reach saturation, not the amplitude, unlike standard viscous damping.
Correspondingly}\,---\,and in contrast to other possible ion-heating mechanisms\,---\,if the fluctuations' amplitude is 
 too small at some time (or in some region of a radially stratified wind) then  energy is not deposited into electrons. Rather,
 the helicity barrier halts the cascade, storing the energy in fluctuations that keep growing, eventually to heat ions through ICWs.
This mechanism is expected to become more robust with decreasing $\beta$, at least for  $\beta\gg m_{e}/m_{i}$ (for $\beta\sim m_{e}/m_{i}$, 
electron heating effects are important at $k_{\perp}\rho_{i}\sim1$). 
Our finding that the helicity barrier occurs even at  modest $\beta$ ($\beta\approx0.3$)  indicates that it should apply nearly everywhere in the
corona. Thus, the ratio of electron to ion heating\,---\,an important input to larger-scale models 
of the solar wind and other astrophysical objects\,---\,is simply $Q_{i}/Q_{e}\approx\varepsilon_{H}/(\varepsilon-\varepsilon_{H})$ for   saturated imbalanced Alfv\'enic turbulence,  nearly independent of $\beta$ for $m_{e}/m_{i}\ll \beta\lesssim1 $. 

A heating ratio that is controlled by imbalance is appealing for understanding the generation of fast- and slow-wind streams
by reflection-driven Alfv\'enic turbulence. 
Fast wind\,---\,which emerges from coronal holes with lower expansion factors \cite{Wang1990} and thus less wave reflection and larger $\varepsilon_{H}/\varepsilon$\,---\,is 
observed to have hotter protons,  strong minor-ion heating, cooler electrons, larger imbalance, and steep transition-range spectra.
Slow wind\,---\,which arises from less ordered fields involving closed loops and/or larger expansion factors, and thus has smaller $\varepsilon_{H}/\varepsilon$ (or
even $\varepsilon_{H}/\varepsilon\ll 1$ in a closed-field region)\,---\,is observed to have cooler protons, little suprathermal minor-ion heating, hotter electrons, 
less imbalance, and does not usually exhibit a steep kinetic transition range.
\revchng{While reflection-driven turbulence models can already  reproduce the observed correlation of expansion factor with wind 
speed based on the radial location of energy deposition\cite{Cranmer2005,Chandran2021}, the additional physics afforded by the helicity barrier yields interesting implications. First, it would 
naturally explain the other aforementioned correlations of electron and ion thermodynamics with wind speed. 
Second, because of electrons' high thermal speeds, a given quantity of energy deposited into ion heat generally drives a 
  higher asymptotic wind velocity than if electrons are heated\cite{Hansteen1995}.
We propose that the helicity barrier could act as a ``switch'' effect, supplementing other turbulent heating physics:
low-expansion regions with a robust barrier would  heat predominantly ions at large radii, 
ideal conditions for generating fast wind speeds;
high-expansion factors would break the barrier, depositing energy into electron heat and exacerbating the inefficient acceleration that 
results from heating at lower radii. 
Thus, by linking plasma thermodynamics
to magnetic-field expansion, it is plausible that the helicity barrier plays an important role in generating the bimodal speed distribution of the solar wind.
}

\setcounter{figure}{0}
\makeatletter 
\renewcommand\figurename{Extended Data Fig.}
\makeatother

\newpage
\section*{Methods}

\subsection*{Hybrid-kinetic simulation method}
The equations of the hybrid-kinetic model solved by \textsc{Pegasus}\texttt{++}  are \cite{Byers1978,Kunz2014a},
\begin{subequations}
\begin{gather}
\frac{\partial f_{i}}{\partial t}+\bm{v} \cdot \grad f_{i}+\left[\frac{e}{m_{i}}\left(\bm{E}+\frac{\bm{v}}{c} \btimes \bm{B}\right)+\frac{\Fu}{m_{i}}\right] \cdot \frac{\partial f_{i}}{\partial \bm{v}}=0,\label{eq: vlasov}\\
\frac{\partial \bm{B}}{\partial t}=-c \grad \btimes (\bm{E} + \Fb) + \eta_{4}\grad^{4}\bm{B},\label{eq: induction}\\
\bm{E}=-\frac{\bm{u}_{{i}} \btimes \bm{B}}{c}+\frac{({\grad} \btimes \bm{B}) \btimes \bm{B}}{4 \pi e n_{{i}}}-\frac{T_{e} \grad n_{{i}}}{e n_{{i}}}.\label{eq: electric}
\end{gather}
\end{subequations}
The kinetic equation \eqref{eq: vlasov} is solved in a six-dimensional $(\bm{x},\bm{v})$ space using a particle-in-cell method, while Faraday's law \eqref{eq: induction} and the kinetic Ohm's law \eqref{eq: electric} are solved on a three-dimensional grid. A single ion species of
charge $e$ and mass $m_{i}$ is assumed, $c$ is the speed of light, and $\Fu$ and $\Fb$ are the forcing terms, which are described belows.  In equation \eqref{eq: electric},  $n_{i}=\int \rmd\bm{v} f_{i}$ and $\bm{u}_{i}=n_{i}^{-1}\int \rmd\bm{v}\,\bm{v}f_{i}$ are computed from $f_{i}$. Equations \eqref{eq: vlasov}--\eqref{eq: electric} are derived from the two-species Vlasov equation by  expanding the electron  equation in $m_{e}/m_{i}\ll1$, assuming quasi-neutrality $n_{e}=n_{i}=n$ and isothermal  electrons (temperature $T_{e}$).          \textsc{Pegasus}\texttt{++} 
uses a second-order accurate predictor-predictor-corrector scheme to enforce the kinetic Ohm's law \cite{Kunz2014a} and has been
highly optimized for  efficient operation on large supercomputing systems. The hyper-resistivity in equation \eqref{eq: induction} is not intended to represent a true 
physical resistivity (it does not contribute to the electric field in the particle push), but is included to dissipate energy in the magnetic field at the smallest grid scales. This is a proxy for electron heating in the model.

\subsubsection*{Simulation parameters}

As discussed in the main text, the simulation domain is elongated by a factor of $6$ ($L_{x}=L_{y}=67.5 d_{i}$ and $L_{z}=404.7d_{i}$, but 
with cubic grid cells, $N_{x}=N_{y}=392$, $N_{z}=2352$) in order to realize  realistic solar-wind conditions near $k_{\perp}\rho_{i}\sim 1$.
These conditions can be  estimated roughly by taking the outer-scale fluctuations to be in approximate critical balance 
$\delta B_{\perp}/B_{0}\simeq L_{\perp}/L_{\parallel}$, and assuming a magnetic spectrum $\mathcal{E}_{B}(k_{\perp})\propto k_{\perp}^{-5/3}$ and 
$k_{\|}\propto k_{\perp}^{2/3}$ in the MHD inertial range \cite{Goldreich1995,Schekochihin2009}. Given the outer  perpendicular scale $k_{\perp0}\rho_{i}\equiv 2\pi\rho_{i}/L_{\perp}\approx0.05$, this suggests that at $k_{\perp}\rho_{i}\sim 1$ the  spectral anisotropy is $k_{\perp}/k_{\|}\simeq \tan86.5^{\circ}$ with fluctuation amplitude $\delta B_{\perp}/B_{0}\simeq 0.06$. This is comparable to observed values  (see Ref.~\citenumns{Chen2016a}, figure 1) justifying the appropriateness 
of our study to solar wind heating.\footnote{In fact, the  observations of Ref.~\citenumns{Chen2016a} show a somewhat \emph{larger} amplitude (less anisotropic) than this estimate; but our simulation also has a larger amplitude, because, due to the helicity barrier,  the  outer-scale amplitude grows noticeably beyond the estimated $\delta B_{\perp}/B_{0}\simeq L_{\perp}/L_{\parallel}$.  }  \revchng{We also note that the  elongated simulation domain implies that the inferred outer-scale\,---\,the scale where $\delta B_{\perp}\sim B_{0}$\,---\,has an extremely 
long turnover time $\simeq (L_{\|}/L_{\perp})^{2}\tau_{A}$ (where $\tau_{\rm A}=L_{\|}/v_{\rm A}$), which is also comparable to the heating
time $\tau_{\rm heat} \sim (3/2m_{i})T_{i}V/\varepsilon$ ($V=L_{\perp}^{2}L_{\|}$  is the volume). Because these outer-scale
timescales exceed the duration of the simulation (${\simeq}18\tau_{\rm A}$), it is apt to consider the outer-scale forcing in our simulation 
as representing a turbulent flux of energy arriving from larger scales, even if the outer scales are decaying\cite{Davidson2004} (as relevant the solar wind). Similarly, we note that the direct effect of solar-wind expansion is negligible at these scales.  A simple estimate can be obtained by matching the simulation's ion-inertial
scale to that in the solar wind,  defining the expansion time as $\tau_{\rm exp}\sim R/U$, where $U$ the bulk solar-wind velocity and $R$ the heliocentric radius. Using parameters similar to PSP's first 
perihelion\cite{Bale2019,Kasper2019}, which also had $\beta_{i}\approx0.3$, yields $\tau_{\rm A} /\tau_{\rm exp} \simeq 8\times 10^{-4} (U/350{\rm kms}^{-1})(R/35R_{\odot})^{-1} (B/80{\rm nT})^{-1}$, 
illustrating that the expansion of the box is negligible ($\simeq 1.5\%$) over the simulation's duration.
}

Other parameters of the imbalanced simulation are chosen to match the balanced turbulence simulation of Ref.~\citenumns{Arzamasskiy2019}, which had a resolution of $N_{\perp}^{2}\times N_{\|}=200^{2}\times1200$, a smaller box with  $k_{\perp0}\rho_{i}\equiv 2\pi\rho_{i}/L_{\perp}\approx0.1$, and saturated amplitude $\delta B_{\perp}/B_{0}\approx 0.14\simeq L_{\perp}/L_{\|}=1/6$.
The energy-injection rate is computed from $\varepsilon/V = C_{A}\delta u_{\perp0}^{2}/\tau_{\rm A}$, with $\delta u_{\perp0}/v_{\rm A}=L_{\perp}/L_{\|}$ and a Kolmogorov constant $C_{A}=0.29$ to match that measured from the  balanced simulation. This 
implies a lower energy injection per unit volume in the imbalanced case, compensating for the slower outer-scale motions in its larger box; stated differently, 
with this $C_{A}$,  turbulence in the larger $k_{\perp0}\rho_{i}\approx0.05$ box would saturate with $\delta B_{\perp}/B_{0}\approx 0.14$ if the forcing were balanced, which
also implies it would have a smaller amplitude (smaller $k_{\|}/k_{\perp}$) at $k_{\perp}\rho_{i}\sim1$. 
The simulation is initialized by randomly drawing particle velocities from a stationary Maxwellian distribution with temperature $T_{i}=T_{e}$.
We use $N_{\rm ppc}=216$ particles per cell and the full-f method \cite{Kunz2014a}, with particles initially evenly distributed within the cell. In computing the gridded  moments of $f_i$, we use two filter passes in order to reduce the impact of the particle noise \cite{Kunz2014a}.

The value of the hyper-resistivity, $\eta_{4}\approx2.4\times10^{-5} d_{i}^{4}\Omega_{i}$,  is also chosen to match that of the balanced simulation (here $\Omega_{i}$ is the ion gyrofrequency and $d_{i}$ is the ion inertial length). Its value is not intended to represent reality, 
but just to absorb magnetic energy that cascades to the grid scales of the simulation.
We have tested the impact of this choice by restarting the simulation in the saturated regime (at $t\simeq 16.3\tau_{\rm A}$) with 
$\eta_{4}\approx1.6\times10^{-5} d_{i}^{4}\Omega_{i}$.  This modification extends  $\mathcal{E}_{B_{\perp}}(k_{\perp})$ to   smaller
scales where it flattens further, as expected, while changing $\varepsilon_{\eta}$ only slightly and making no  noticeable difference to the diagnostics presented in 
 Figs.~3 and 4 of the main text. We 
are thus confident that the chosen $\eta_{4}$ is appropriate and that  $\varepsilon_{\eta}$ is not adversely affected by grid-scale effects.

\subsubsection*{Forcing}

The plasma is driven at the largest scales in the box with the forcing terms $\Fu$ and $\Fb$ in equations \eqref{eq: vlasov} and \eqref{eq: induction}, respectively.
Given that our elongated box is designed to represent a small patch of a 
much larger system, these  terms are supposed to mimic crudely the effect of the larger-scale turbulence on our box's outer scale. As in Ref.~\citenumns{Meyrand2021}, we opt 
to design the forcing to inject energy and cross helicity at a constant rate in time, which necessarily requires adjusting  $\Fu$ and $\Fb$ to respond 
to the state of the plasma.
We thus define $\Fu=f^{U} \bb{F}_{0}$ and $\grad\btimes \Fb=f^{B} \bb{F}_{0}$, where the forcing function $\bb{F}_{0}$ is divergence-free and purely 
perpendicular to $\bb{B}_{0}$ (no $z$ component), which implies the forcing is nearly purely Alfv\'enic (its compressive part is small). $\bm{F}_{0}$  is evolved in time as an Ornstein-Uhlenbeck process with  correlation time $t_{\rm corr}=\tau_{\rm A}/2$ for each mode with $2\pi/L_{j}\leq k_{j}\leq 4\pi/L_{j}$, where $j$ represents each direction $x$, $y$, and $z$. At each time step, we compute $n\bm{u}_{i}\bcdot\Fu$ and
$\bm{B}\bcdot(\grad\btimes\Fb)$ and adjust the values of $f^{U}$ and $f^{B}$ so as to make the injected energy and cross helicity  equal to their desired values ($\varepsilon$ and $\varepsilon_{H}$, respectively,
with $\varepsilon_{H}=0.9\varepsilon$). This process requires inverting the curl to compute $\Fb$, which is carried out using a Fourier transform; by adding
the magnetic force in this way and evolving $\bm{B}$ using the standard constrained-transport algorithm of \textsc{Pegasus}\texttt{++}, we ensure $\grad\bcdot\bm{B}=0$ to machine precision.

This forcing method is an extension (to allow for  imbalance) of the default routines implemented in the  \textsc{Athena} code \cite{Stone2008}, and has thus been used 
in a number of previous works (e.g., Ref.~\citenumns{Lynn2012}). We have tested it by  measuring separately the energy and cross-helicity injection, which agree almost 
perfectly with  the input values, and by testing the full energy budget of the simulation (see below).
A possible downside of the method is that the forcing normalization, $f^{U}$ and $f^{B}$, can change more rapidly 
than the spatial form of the force $\bm{F}_{0}$. In the simulation, we see occasional sudden changes in $f^{U}$ and $f^{B}$ that seem to be caused by the 
plasma flow and magnetic-field perturbations (which dominantly propagate in the $-\ez$ direction) moving out of phase with the large-scale spatial 
structure  of the forcing (determined by the slow evolution of $\bm{F}_{0}$). 
In order to assess the impact of this effect, we restarted the simulation at $t=10\tau_{\rm A}$ with a modified version of the forcing that 
limited the change in  $f^{U}$ and $f^{B}$ across a timestep $\delta t$ to $\exp(\pm \Omega_{i} \delta t/10 )$, where the $\pm$ was decided based on whether
 $f^{U}$ and $f^{B}$ were above or below the optimal value that gave the input energy and cross-helicity injection. This reduced the sudden changes (high-frequency power) in $f^{U}$ and $f^{B}$ at the 
 cost of causing $\varepsilon(t) $ and $\varepsilon_{H}(t) $ to vary 
 significantly (by around $\pm40\%$) in time. This change, however, made no noticeable difference  to the heating, distribution 
 function, and  spectra, lending us confidence in the robustness of our results.
 Finally, we have confirmed that the helicity barrier also forms robustly in the reduced model of Ref.~\citenumns{Meyrand2021} when the outer
scales are forced with white-in-time noise (as opposed to with constant energy and cross-helicity injection). Thus, we do not expect our results to be particularly sensitive to the design of the forcing.

\begin{figure}
\centering
\includegraphics[width=0.75\columnwidth]{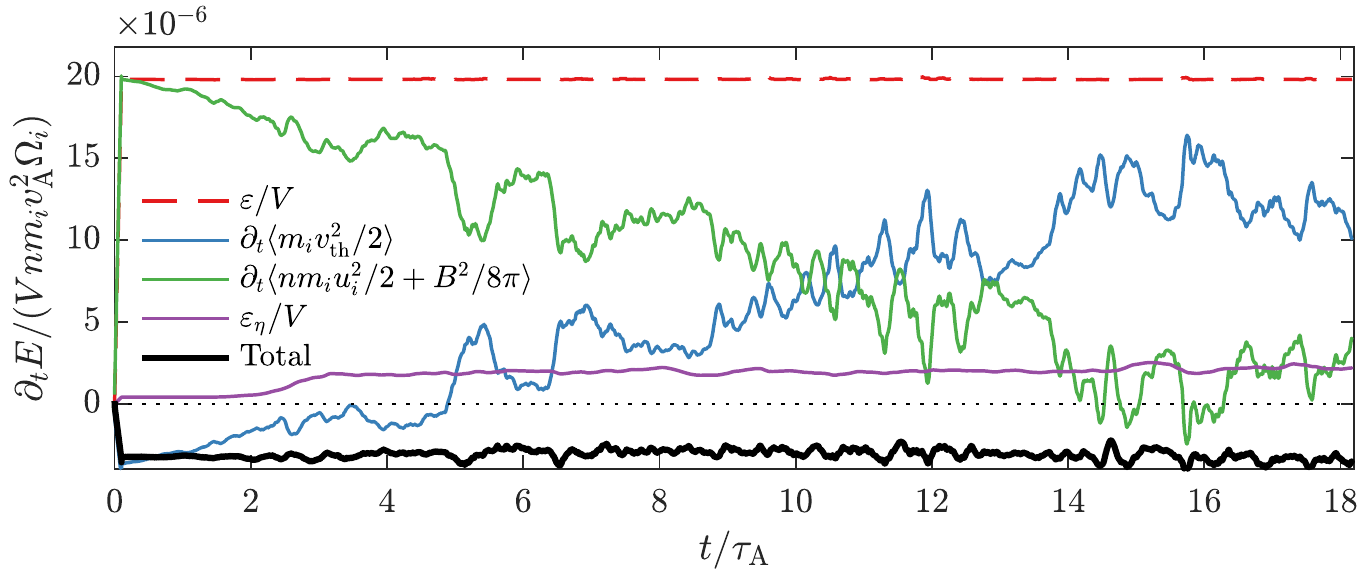}
\caption{\textbf{Electric field noise and numerical cooling.} Contributions to the energy budget per unit volume of the imbalanced simulation from the energy injection ($\varepsilon/V$; dashed red line), increase in thermal energy $Q_{i}= \partial_t\langle m_i v_{\rm th}^2/2\rangle $ (blue line), growth 
rate of mechanical energy $\partial_t\langle nm_i u_i^2/2+B^2/8\pi\rangle $ (green line), and resistive dissipation $\varepsilon_{\eta}/V$ ($V$ is the volume and $\langle\,\dots\,\rangle$ denotes a box average).
The black line shows the total energy budget ${\rm Total}= \varepsilon/V - \varepsilon_{\eta}/V -  \partial_t\langle m_i v_{\rm th}^2/2\rangle - \partial_t\langle nm_i u_i^2/2+B^2/8\pi\rangle $, which is constant and negative, indicating numerical cooling that is effectively independent of the turbulence or the heating of ions. }
\label{fig: energy budget}
\end{figure}
\begin{figure}
\centering
\includegraphics[width=0.7\columnwidth]{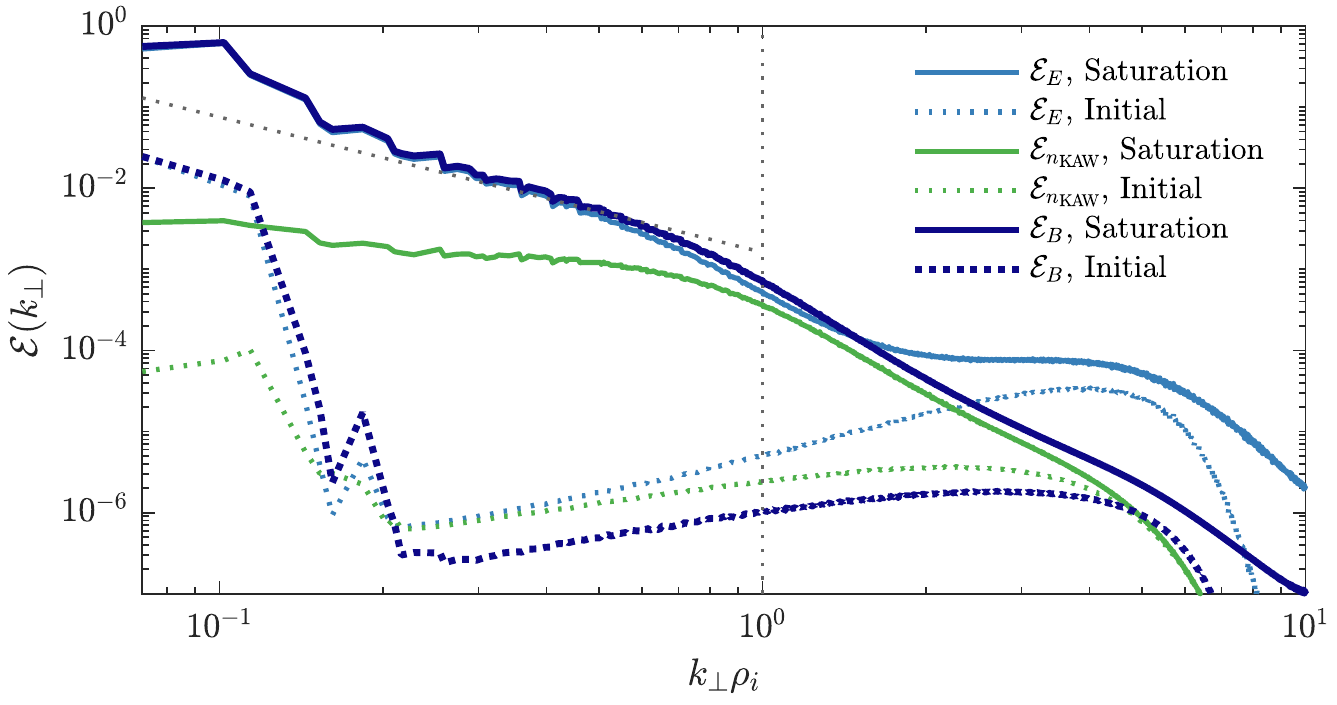}
\caption{\textbf{The effect of particle noise on turbulence spectra.} Perpendicular ($k_{\perp}$) spectra of the magnetic field ($\mathcal{E}_{\bm{B}}$), electric field $\mathcal{E}_{\bm{E}}$, and KAW-normalized density $\mathcal{E}_{\nkaw}=\beta_{i}(1+2\beta_{i})\mathcal{E}_{n}$ in the saturated state (solid lines) and at very early times (averaged over $t\leq0.2\tau_{\rm A}$). The latter is from before the turbulence has developed and is thus a proxy for the noise floor in a given quantity.  At the smallest scales, $k_{\perp}\rho_{i}\gtrsim 3$, spectra are only modestly above the noise floor and therefore
uncertain. }
\label{fig: energy budget}
\end{figure}

\subsubsection*{Electric-field noise and numerical cooling}\label{sub: noise and energy }

A persistent problem for the particle-in-cell method is the influence of electric-field noise, which arises due to
random density noise from the finite number of numerical particles. In the full-f hybrid-kinetic method with the \textsc{Pegasus}\texttt{++}
algorithm,
a key impact of this noise is a numerical cooling, which slowly drains thermal energy from the system.
This was removed from Fig.~1 of the main text for clarity, but can be accurately assessed by
computing the thermal energy budget, which is shown in Extended Data Fig.~1. The sum of the various 
contributions to the total rate of change of energy, shown by the thick black line, would equal zero if energy were conserved, but is instead negative,
indicating numerical cooling. This feature\,---\,although clearly undesirable\,---\,can be avoided only by increasing the number 
of particles per cell (thus increasing computational cost), or by using more  filter passes  
(thus decreasing the effective dynamic range). Its nearly constant value throughout the simulation\,---\,including 
at very early times $t\lesssim \tau_{\rm A}$, when the forcing is fully absorbed by large-scale mechanical energy and there is
no resistive dissipation\,---\,suggests its properties
are mostly separate from the development of turbulent heating in the system, justifying tolerating its presence and removing it from our energy estimates. 

Another effect of the particle noise is its direct influence on the spectra. Noisy fluctuations cause an artificial bump in all quantities near the
grid scale, which accounts for some  of the small-scale flattening of the electric- and magnetic-field spectra in Fig.~2 of the main text.
This can be quantified by computing spectra from the very early stages of the simulation, before the turbulence has developed, 
when the small-scale fluctuations are purely a result of particle noise. We compare these to the saturated-turbulence spectra  in Extended Data Fig.~2. The general conclusion is  
 that for $k_{\perp}\rho_{i}\gtrsim3$, noise makes a reasonable contribution to the spectra, which are thus uncertain in this range.
This effect is more severe in this imbalanced simulation than other previous turbulence simulations with \textsc{Pegasus}\texttt{++} \cite{Arzamasskiy2019,Cerri2021}
precisely because of the transition range, which makes the sub-$\rho_{i}$ fluctuations very small in magnitude. 
Again, it can only be ameliorated by increasing the number of particles per cell, which yields small gains for large computational expense (the noise scales ${\propto}N_{\rm ppc}^{1/2}$). 
If one subtracts the noise spectra from the saturated spectra\,---\,likely a reasonable procedure because the  noise power is dominated by modes of high $k_{\perp}$ and $k_{z}$,  while the majority of high-$k_{\perp}$  power at saturation resides at lower $k_{z}$\,---\,our main conclusions still hold, with $\mathcal{E}_{\bm{B}}$ and $\mathcal{E}_{\bm{E}}$ re-flattening at $k_{\perp}\rho_{i}\simeq 2$ and a ${\sim}k_{\perp}^{-0.8}$ KAW range in $\mathcal{E}_{\bm{E}}$ (not shown).

Finally, we note
that a different preliminary simulation with three-times larger $\varepsilon$,  four filter passes, and $N_{\rm ppc}=128$
reproduced the key early time features discussed above and in the main text, including a double-kinked electric-field spectrum (this simulation was run only until $t\simeq 5\tau_{\rm A}$, which is  before magnetic-field spectrum develops a double kink). \revchng{In addition to causing the large-scale energy to grow faster in time}, the 
larger $\varepsilon$ in this case caused the noise in spectra and the numerical cooling to be proportionally 
much smaller. We are thus confident that the physical features reported in the main text are robust.

\subsection*{Measurement of the parallel spectrum}

In Figs.~2 and 3 of the main text, we measure parallel spectra using a field-line following method, which we describe here.
While structure-function methods are more commonly used  to study anisotropic MHD turbulence \cite{Schekochihin2020}, 
the extremely steep parallel spectra (up to $\sim k_{\|}^{-6}$) caused by 
helicity barrier are not well captured by structure functions \cite{Cho2009}. 
Our field-line-following method computes a ($k_{\perp}$, $k_{\|}$) spectrum by first constructing $N_{\rm lines}$ magnetic-field lines 
by solving $\rmd\bm{r}/\rmd s = \eb(\bm{r})=\bm{B}/|\bm{B}|$ from $s=0$ to $s=L_{\rm lines}$, 
where the periodicity of the system allows $L_{\rm lines}>L_{\|}$ if desired. The field of interest $X(\bm{x})$
(e.g., $X=B_{y}$ or $X=E_{z}$) is then Fourier-filtered to a given bin in $k_{\perp}=\sqrt{k_{x}^{2}+k_{y}^{2}}$, giving $X_{k_{\perp}}(\bm{x})$,
which is then
interpolated onto the coordinates $\bm{r}(s)$ of the previously computed field lines.  One then computes the spectrum of $X_{k_{\perp}}[\bm{r}(s)]$ in the field-line ($s$) direction, which becomes a single $k_{\perp}$ slice 
of $\mathcal{E}_{X}(k_{\perp},k_{\|})$. Repeating this process across a grid of $k_{\perp}$ yields the full 2D spectrum.\footnote{If this
process is applied to compute a $(k_{\perp},k_{z})$ spectrum [i.e., $\bm{r}(s)$ in the $\ez$ direction], it yields the same 
result as computing the spectrum in the standard way with a 3D Fourier transform.} Through experimentation, 
we found $L_{\rm lines}=10L_{\|}$ and $N_{\rm lines} = N_{x}N_{y}L_{\|}/L_{\rm lines}$ gives high-quality results (though results are almost independent of 
$L_{\rm lines}$  for $L_{\rm lines}\gtrsim 2L_{\|}$).
In addition, a Hamming filter is used to compute the $k_{\|}$ spectra because the field-line ($s$)  direction is non-periodic. Because the field-perpendicular plane is assumed to 
 be the $x,y$ plane, the method is valid in the ``reduced-MHD'' limit of $L_{\perp}/L_{\|}\sim \delta B_{\perp}/ B_{0}\ll1$.
 A pure parallel spectrum, as in the inset of Fig.~2 of the main text, is computed from  $\int\rmd k_{\perp}\,\mathcal{E}_{X}(k_{\perp},k_{\|})$, which 
 recovers the standard field-line parallel spectrum \cite{Meyrand2019}.
\begin{figure}
\centering
\includegraphics[width=0.55\columnwidth]{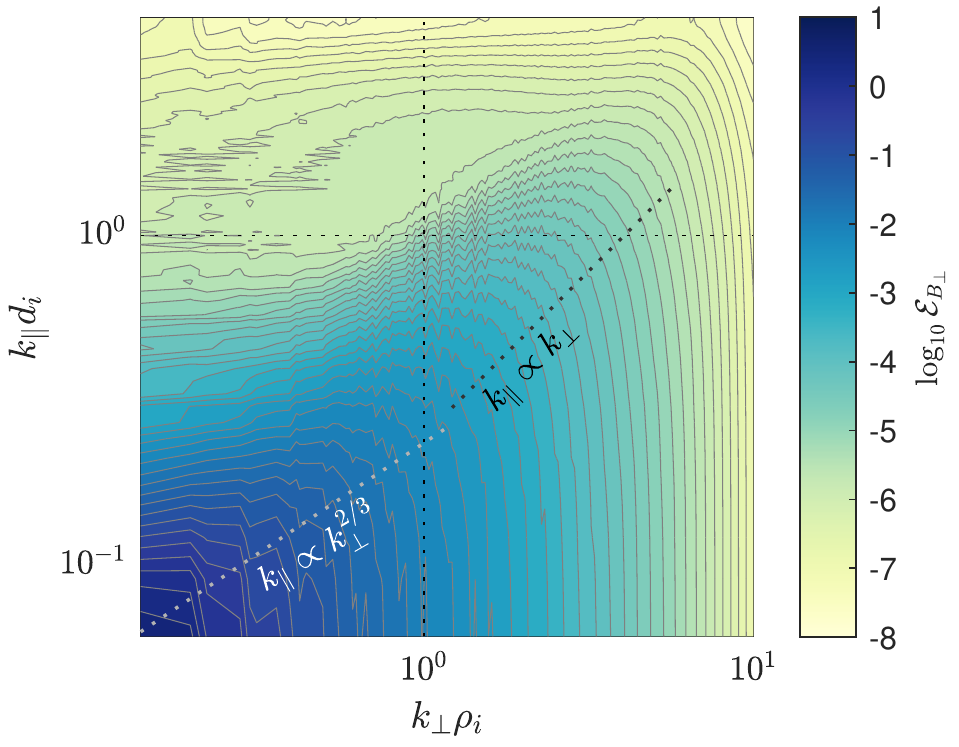}
\caption{\textbf{Measurement of the parallel spectrum. }Two-dimensional perpendicular magnetic-field spectrum $\mathcal{E}_{B_{\perp}}(k_{\perp},k_{\|})=\mathcal{E}_{B_x}(k_{\perp},k_{\|})+\mathcal{E}_{B_y}(k_{\perp},k_{\|})$
from the balanced simulation. The method recovers the large- and small-scale scalings of $k_{\|}$ with $k_{\perp}$
measured using structure functions  (Ref.~\citenumns{Arzamasskiy2019} figure 2), as well as the predicted 2D  spectrum in the $k_{\perp}\rho_{i}<1$ range (see Ref.~\citenumns{Schekochihin2020}, appendix B).   }
\label{fig: lev 2d spec }
\end{figure}

We have compared our method to the more conventional method of computing 2D spectra via structure functions, finding reasonable agreement where both are appropriate. 
Extended Data Fig.~3 shows  $\mathcal{E}_{B_{\perp}}(k_{\perp},k_{\|})=\mathcal{E}_{B_{x}}+\mathcal{E}_{B_{y}}$ for the balanced simulation,
illustrating how we recovered the previously discussed $k_{\|}\sim k_{\perp}^{2/3}$ and $k_{\|}\sim k_{\perp}$ ranges above and 
below the ion-Larmor scale $k_{\perp}\rho_{i}\sim 1$ \cite{Arzamasskiy2019}. Our method also recovers  the expected extremely steep scaling in $k_{\|}$ ($\mathcal{E}_{B_{\perp}}\propto k_{\|}^{-6}$) predicted by theory at low $k_{\perp}$ (see appendix B of \citenumns{Schekochihin2020}; not shown).
A more quantitative comparison reveals that the most notable difference compared to the structure-function method is that the 2D spectrum
is shifted upwards somewhat (to larger $k_{\|}$); thus, compared to the structure-function method, the field-line-following method generally  estimates $k_{\|}(k_{\perp}) $ to be a  factor of $\simeq2$ larger across all $k_{\perp}$.  

Note that the helicity spectrum (Fig.~3C in the main text) was computed by binning in $(k_{\perp},k_{z})$ rather than $(k_{\perp},k_{\|})$, which
 gave a cleaner spectrum. The difficulty with $(k_{\perp},k_{\|})$ seems to  relate to $\mathcal{H}_{\bm{k}}$ explicitly involving $\bm{k}$ in its definition \cite{Matthaeus1982,Howes2010}, which may cause
problems with  the non-orthogonal 
coordinate system $(k_{x},k_{y},k_{\|})$. The  $(k_{\perp},k_{z})$ spectrum of $B_{\perp}$ looks broadly similar to its $(k_{\perp},k_{\|})$ spectrum, particularly in the $k_{z}>k_{\perp}$
region of ICWs, so Fig.~3B and 3C of the main text can be usefully compared on a qualitative level. 

\subsection*{Heating through quasi-linear diffusion on ICWs}

\subsubsection*{Computation of the resonance contours for parallel ICWs}

Here we describe in more detail the computation of the quasi-linear (QL) ICW ``scattering contours'' shown in Fig.~4 of the main text. 
QL theory \cite{Kennel1966} is based on the idea that particles interact strongly with a spectrum of ICWs if
they satisfy the cylotron-resonance condition \begin{equation}
\omega_{\bm{k}}-k_{\|}w_{\|}=\pm\Omega_{i}.\label{eq: resonance}
\end{equation}
Here $\omega_{\bm{k}}$ is the wave's frequency  and
$w_{\|}$ is the ion's parallel velocity (both are measured in the fluid frame in which the plasma is stationary); equation \eqref{eq: resonance} expresses the condition for the Doppler-shifted wave frequency to be resonant with the ion's gyromotion, which causes the wave and the particle to interact strongly.
In the frame moving with the wave's phase velocity ${v}_{\rm ph}=\omega_{\bm{k}}/k_{\|}$,
the magnetic-field perturbation is constant in time. This means that the  electric field must be purely potential in this frame ($\bm{E}=-\grad\Phi$), and
a particle should approximately conserve its kinetic energy as it scatters from the wave. Thus, if Eq.~\eqref{eq: resonance} has only one solution for each $w_{\|}$\,---\,in other words, when there  exist only waves of a single $k_{\perp}$ (e.g., if only parallel waves exist) or when $\omega_{\bm{k}}$ is a function of $k_{\|}$ only \,---\,particles diffuse only along specific ``scattering contours'' in the $w_{\perp}$--$w_{\|}$ plane. If the
waves are non-dispersive (i.e., if $\omega_{\bm{k}}\propto k_{\|}$) these are semi-circles of constant particle energy in the wave's frame $\ekwave= (w_{\|}-\omega_{\bm{k}}/k_{\|})^{2}+ w_{\perp}^{2}$. More generally, the scattering contours are defined by the null solutions $\eta(w_{\perp},w_{\|})={\rm const.}$ of the QL diffusion operator  \cite{Kennel1966,Chandran2010a,Isenberg2011},
\begin{equation}
\left[1-\frac{w_{\|}}{v_{\mathrm{ph}}(w_{\|})}\right] \pD{w_\perp}{\eta} + \frac{w_{\perp}}{v_{\rm ph}(w_{\|})} \pD{w_\parallel}{\eta} =0,\label{eq: contour equation}
\end{equation}
where $v_{\rm ph}(w_{\|})=\omega_{\bm{k}}/k_{\|}$ for the $w_{\|}$ that satisfies the resonant condition \eqref{eq: resonance}.
If $f_{i}$ is a decreasing function   along the scattering contours, the QL diffusion
heats the plasma in the range of $w_{\|}$ where there is appreciable  power in  waves of the corresponding resonant $k_{\|}$; if $f_{i}$ is an increasing function along the scattering contours, the distribution function will become unstable, growing the wave power in the resonant range of $\bm{k}$.

In order to compute the scattering contours, we assume the cold-plasma ICW dispersion relation for parallel propagating waves ($k_{\perp}=0$), 
which  is \begin{equation}
\omega_{\bm{k}} = k_{\|}v_{\rm A} \sqrt{1-\omega_{\bm{k}}/\Omega_{i}} \quad \Rightarrow \quad v_{\rm ph}(k_{\|}) = \frac{v_{\rm A}}{2\Omega_{i}}\left( \sqrt{k_{\|}^{2} v_{\rm A}^{2}+ 4\Omega_{i}^{2}}-k_{\|}v_{\rm A}\right).\label{eq: ICW DR}
\end{equation}
A parametric solution for $\eta$ can be found by solving Eqs.~\eqref{eq: resonance} and \eqref{eq: ICW DR} for $v_{\rm ph}(w_{\|})$, and then using this in the solution of  Eq.~\eqref{eq: contour equation}, $w_{\perp}^{2}+w_{\|}^{2}-2\int\rmd w_{\|}'\,v_{\rm ph}(w_{\|}')={\rm const} $.
This yields the contours \cite{Isenberg1996} 
\begin{equation}
\frac{w_{\perp}^{2}}{v_{\rm A}^{2}}+\frac{1}{y^{2}} +\ln y-\sinh^{-1} \frac{y}{2} = {\rm const},\label{eq: resonance contours}
\end{equation}
where $y = k_{\|}v_{\rm A}/\Omega_{i}$ relates to $w_{\|}$ implicitly through Eq.~\eqref{eq: resonance}  (the full explicit solution is uninformatively complex). 

The QL diffusion process relies on maintaining wave power 
in the relevant range of $k_{\|}$, but the simulation exhibits a steep drop in the spectrum for  $k_{\|}d_{i}\gtrsim1$ (see Fig.~3 of the main text).\footnote{This is likely because linear ICWs become strongly damped at $k_{\|}d_{i}\simeq1$ at $\beta_{i}=0.3$ (as can be shown by solving the
hot-plasma dispersion relation).}
This implies a minimum $|w_{\|}|=w_{\|,d_{i}}$ above which $f_{i}$ should tend to flatten along the scattering contours and below which it should not.  This is computed from Eqs.~\eqref{eq: ICW DR} and~\eqref{eq: resonance} by solving for $w_{\|}$ at $k_{\|}=d_{i}^{-1}$, giving  \begin{equation}
w_{\|,d_{i}}\approx-\frac{1}{2}\left(3-\sqrt{5}\right)v_{\rm A}.\label{eq: wld}
\end{equation}
The contours \eqref{eq: resonance contours} and the cutoff \eqref{eq: wld},  which are plotted in Fig.~4A of the main text, provide an exceptionally good match to the shape of  $f_{i}$.

\subsubsection*{Oblique ICWs}\label{subsub: oblique ICW}

The above calculation  assumes that only parallel waves exist in the plasma. In the presence of oblique waves, the dependence of $v_{\rm ph}(k_{\|}) $ 
on $k_{\perp}$ means that a range of scattering contours exists for a  given $w_{\|}$. Refs.~\citenumns{Chandran2010a} and \citenumns{Pongkitiwanichakul2010} provide
a compelling argument as to why it is the parallel ICW scattering contours that should determine the form of $f_{i}$, 
even if oblique modes provide the primary heating power  (see also Ref.~\citenumns{Isenberg2011}).  They note that the dependence of $v_{\rm ph}(k_{\|}) $ on~$k_{\perp}$ is 
such that higher-$k_{\perp}$ modes produce scattering contours that are steeper [i.e., $(\partial\eta/\partial w_{\perp})/(\partial\eta/\partial w_{\|})$ is larger]. A QL diffusion process along these contours will thus 
produce an $f_{i}$ that is an increasing function along the resonance contours of parallel modes [Eq.~\eqref{eq: contour equation}], which 
will  be unstable and generate parallel ICWs. A spectrum of driven oblique modes will therefore generate parallel ICWs in the process
of heating\,---\,effectively a kinetic mechanism for spectral transfer from oblique to parallel modes\,---\,creating an $f_{i}$ that is almost flat along the \emph{parallel} ICW resonance contours. 

This 
phenomenology  provides a reasonable explanation for the behavior that we observe in our simulation: in the saturated stage, there appears a significant population 
of parallel ($k_{\perp}\approx 0$) ICWs (see main text Fig.~3B), even though the turbulent cascade of  energy to such modes is likely 
slow compared to the power input into oblique modes (indeed, the power in parallel modes is quite 
small earlier in the simulation).\footnote{A careful examination shows that the contours of $f_{i}$ are
very slightly steeper than the parallel-ICW scattering contours, which should also be expected from this phenomenology:
the shape of $f_{i}$ results from a balance between  heating from oblique modes (which occurs on contours that are considerably steeper than 
those of $f_{i}$) and cooling/instability from parallel modes. However, we caution that this difference could easily have other 
explanations: e.g.,  differences between the $\beta_{i}=0.3$ and cold-plasma ICW dispersion relations or higher-order resonances. }
This also suggests that the ICW heating process can continue even after  $f_{i}$ becomes perfectly flat along the parallel ICW resonance contours. \revchng{This is an important feature of this heating process for application to the solar wind }and
 seems to be what we observe in the saturated state: after $t\approx14\tau_{\rm A}$ heating  continues but $f_{i}$ expands outwards across the scattering contours. Potential complications with this scenario arise from higher-order resonances (which are possible with oblique but not with parallel ICWs) and electron damping by ICWs at $k_{\|}d_{i}\ll1$ \cite{Gary2004a}, but these are generally expected to be unimportant
to the overall physics \cite{Chandran2010a}.

\revchng{Finally, we emphasize that by the end of the simulation, the average thermal anisotropy of the full distribution is only $T_{\perp}/T_{\|}\approx 1.03$. This value is small for two main reasons: first, because $\tau_{\rm heat}\simeq 54 \tau_{\rm A}$ is longer than the simulation duration and second, because
of the development of the beam, which contributes to the parallel temperature. Given that this $T_{\perp}/T_{\|}$ is far below the usual bi-Maxwellian ICW instability threshold of $T_{\perp}/T_{\|}\approx 1.7$ at $\beta\approx0.3$ that is often used in observational studies\cite{Hellinger2006}, it is clear that the 
detailed shape of the distribution function must be considered in order to understand its stability to parallel ICWs and other wave modes.}

\subsubsection*{Computation of $D_{\perp\perp}^{E}$}\label{subsub: Dpp comp}

The perpendicular energy diffusion coefficient provides a useful quantitative diagnostic of the plasma heating mechanism. It
 is defined by assuming that 
 \begin{equation}
 \pD{t}{f_i} = \pD{e_\perp}{} \left( D^E_{\perp\perp} \pD{e_\perp}{f_i}\right),\label{eq: Dprpprp defn}
\end{equation}
where $e_{\perp}=w_{\perp}^{2}/2$ is the perpendicular kinetic energy per unit mass. The assumption that the evolution of $f_{i}$ is described
by \eqref{eq: Dprpprp defn}, which is simply a diffusion equation
in perpendicular energy, hinges on perpendicular
heating dominating over parallel heating, as is indeed the case in our simulation (at least in the $w_{\|}\lesssim 0$ part of velocity space). 
In standard QL theory (ignoring the oblique--parallel mode interaction discussed above), the resonance condition
depends only on the parallel velocity and the fluctuation spectrum, implying that the velocity diffusion coefficient is independent of $w_{\perp}$ and thus 
 that the energy diffusion coefficient scales as $D_{\perp\perp}^{E}\propto w_{\perp}^{2}$.
 Numerically, we compute $D_{\perp\perp}^{E}$ directly from Eq.~\eqref{eq: Dprpprp defn} using 
\begin{equation}
D_{\perp\perp}^{E} = \left(\pD{e_\perp}{f_i}\right)^{-1}\int_{0}^{e_{\perp}}\rmd e_{\perp}' \, \pD{t}{f_i} ,\label{eq: Dprpprp compute}
\end{equation}
with the full expression integrated over a range of $w_{\|}$ (this is method II of Ref.~\citenumns{Vasquez2020}). We have also compared this result to a direct measurement of the perpendicular heating from $ D_{\perp\perp}^{E}= -(\partial Q_{\perp}/\partial e_{\perp}) / ({\partial f_{i}}/{\partial e_{\perp}})$, where $ \partial Q_{\perp}/\partial e_{\perp}$ is a direct measure of the heating of particles by electric fields based on $ \partial Q_{\perp}/\partial w_{\perp} = e \langle \bm{w}_{\perp}\bcdot\bm{E}_{\perp} f_{i}\rangle$, and $\langle \bm{w}_{\perp}\bcdot\bm{E}_{\perp} f_{i}\rangle$ is 
computed  from ion velocities and electric fields during the simulation. This measurement, although noisier, recovers very 
similar results to Eq.~\eqref{eq: Dprpprp compute}.
Further discussion can be found in Refs.~\citenumns{Vasquez2020} and \citenumns{Cerri2021}, in particular in appendix A of Ref.~\citenumns{Cerri2021}.

\subsection*{Stochastic heating}

\begin{figure}
\centering
\includegraphics[width=0.65\columnwidth]{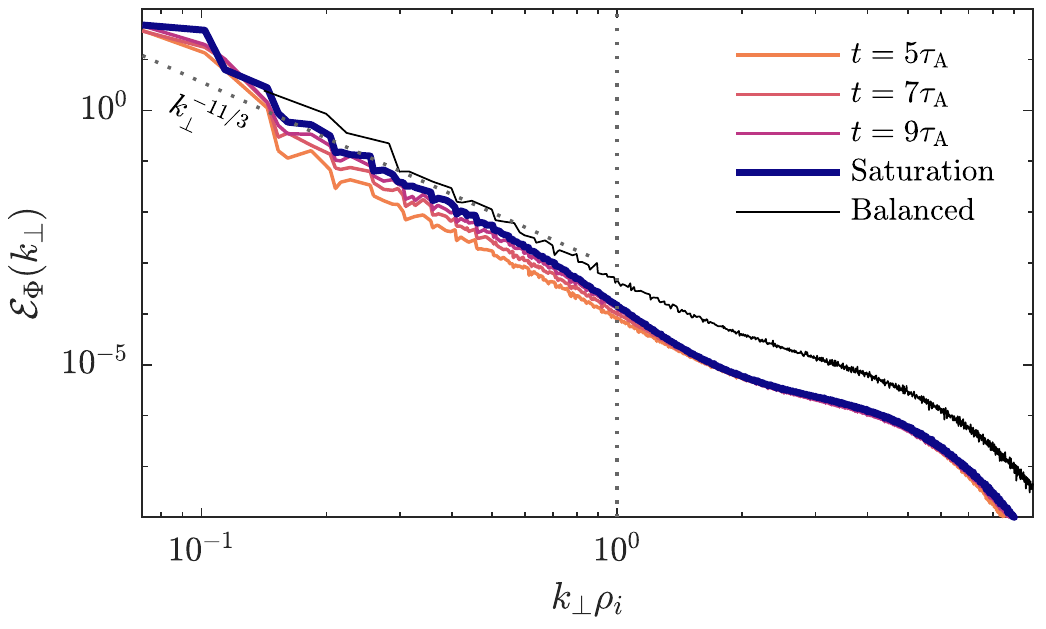}
\caption{\textbf{Assessment of the influence of stochastic heating.} We show perpendicular spectra of the electric potential $\Phi$, computed from the curl free part of $\bb{E}$. Colored lines show various 
times from the imbalanced simulation. The black line shows the  equivalent balanced simulation, which is  averaged over the early period of the simulation (between $t=3.5\tau_{\rm A}$ and $t=4.5\tau_{\rm A}$) when stochastic-ion heating absorbs the majority of the turbulent energy flux \cite{Arzamasskiy2019}. Despite the 
larger turbulence amplitude in the imbalanced simulation, the electric-potential fluctuations around $k_{\perp}\rho_{i}\sim 1$\,---\,those 
important for stochastic heating\,---\,are smaller. }
\label{fig: phi}
\end{figure}

Here we address  whether, in addition to quasi-linear ICW heating, stochastic  ion heating \cite{Chandran2010}  
might play an important role in turbulence with a helicity barrier. The mechanism 
is of particular interest, given its prominence in previous theoretical and observational studies \cite{Klein2016,Martinovic2020,Cerri2021}.
Possible scenarios  could involve multiple heating mechanisms operating at a particular
time, or a transition from one heating mechanism to another as $f_{i}$ changes shape in time.
We find no evidence for such behavior\revchng{ in this simulation}: $D_{\perp\perp}^{E}$ appears to maintain its QL scaling $D_{\perp\perp}^{E}\propto w_{\perp}^{2}$ throughout, 
and $f_{i}$ does not deviate from the scattering contours as it evolves in the saturated state.

We speculate that in this simulation, 
the lack of stochastic heating is simply a consequence of its small electric-potential ($\Phi$) fluctuations  around $k_{\perp}\rho_{i}\sim 1$, 
which are required in order to make ion gyro-orbits sufficiently random to cause heating \cite{Chandran2010}. In Extended Data Fig.~4,
we compare the spectrum of $\Phi$ ($\mathcal{E}_{\Phi}$) in the imbalanced simulation and the balanced simulation 
\cite{Arzamasskiy2019}. $\mathcal{E}_{\Phi}(k_{\perp}\rho_{i}=1)$  grows only modestly during the imbalanced simulation, 
despite the growth of $\mathcal{E}_{\Phi}$ at larger scales, because of the steep drop at $k_{\perp}\rho_{i}<1$ due to the helicity barrier. Coupled with its larger 
box, we see that even though the imbalanced simulation saturates with larger amplitude turbulence, its ion-gyroscale $ \Phi$ fluctuations
are smaller than in the balanced run. Given that stochastic heating plays only a modest role in this
 balanced run\,---\,it becomes subdominant after several turnover times due to flattening of the core of $f_{i}$ \cite{Arzamasskiy2019,Cerri2021}\,---\,this difference in $\Phi$ may
be sufficient to render stochastic heating unimportant in imbalanced turbulence at these parameters.

It is unclear whether stochastic heating will play a role in other regimes or  over longer timescales. 
So long as gyroscale fluctuations have sufficient amplitudes, stochastic heating is expected to be more robust at lower $\beta$ because more heating occurs before it is quenched by the flattening of the core of $f_{i}$ \cite{Chandran2010}. This conclusion supported by the $\beta\approx1/9$ hybrid 
simulation of Ref.~\citenumns{Cerri2021}. \revchng{On the other hand, test-particle simulations show a strong reduction in the efficiency of  heating
of $w\approx v_{\rm A}$ ($\beta\approx1$) particles  in imbalanced, compared to balanced, turbulence \cite{Teaca2014}. If a similar  reduction  occurs also in the $w\ll v_{\rm A}$ low-$\beta$ regime, the effectiveness of stochastic heating in turbulence with  a helicity barrier may also be limited. }Further work is needed. However, it is worth noting that even if stochastic heating, rather than ICW heating, eventually absorbs the turbulent
energy flux, the helicity barrier could remain a key ingredient in solar-wind heating. Just like for ICW heating, the barrier would allow turbulent fluctuations to grow  their amplitude sufficiently to enable ion heating, rather than fluctuations dissipating their energy into electron heating if their amplitude is initially too small.

\subsection*{Landau damping and the ion beam}

An interesting feature of the ion distribution function shown in Fig.~4 of the main text is the plateau $w_{\perp}\approx0$, $w_{\|}\sim v_{\rm A}$. This
forms a modestly super-Alfv\'enic beam with similar properties \cite{Marsch1982,Marsch2006} and directionality \cite{He2015a} to those observed  in the solar wind. Here we 
present evidence that this feature is a result of Landau damping of perpendicular Alfv\'en waves as they become dispersive (speed up) near $k_{\perp}\rho_{i}\sim1$. 
Test-particle calculations have shown this process to be highly effective \cite{Li2010}. 
\begin{figure}
\centering
\includegraphics[width=0.7\columnwidth]{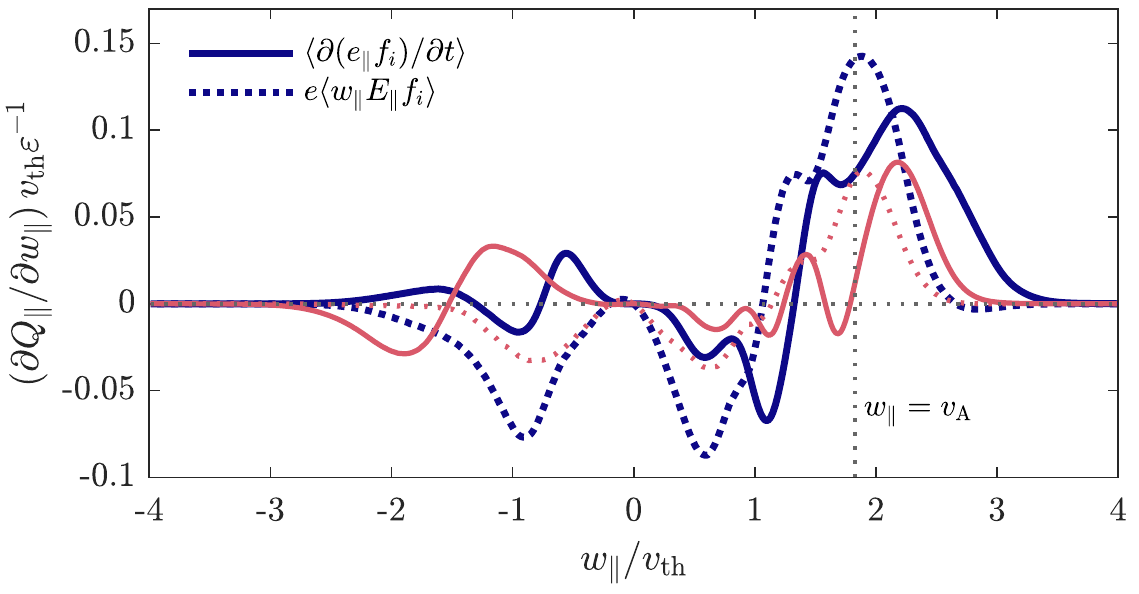}
\caption{\textbf{Development of the ion beam.} We compare the rate of change parallel thermal energy (solid lines; see text) with the  work done on particles by the parallel electric field $e\langle w_{\|}E_{\|}f_{i}\rangle$ (dotted lines). The thick dark-blue lines show the saturated state and the orange-pink lines show $t=7\tau_{\rm A}$. The 
similarity of the magnitude and general shape of the two measures of heating suggests that Landau damping is responsible for the formation of the ion beam. }
\label{fig: beam}
\end{figure}

Extended Data Fig.~5 compares the  measured parallel heating of the distribution function,
\begin{equation}
\left\langle \pD{t}{e_{\|}f_{i}}\right \rangle (w_{\|})= \int \rmd e_{\perp}\,\frac{1}{2}w_{\|}^{2}  \pD{t}{f_{i}},
\end{equation}
to the parallel heating inferred from the work done by the parallel electric field $e\langle w_{\|}E_{\|}f_{i}\rangle$, which 
is computed during the simulation from particle trajectories. We 
see a clear peak in both quantities at $w_{\|}\approx v_{\rm A}$.
As discussed in  Methods (see equation 8 therein), both methods measure plasma heating but they can differ by a total derivative. 
Their similarity in Extended Data Fig.~5\,---\,in particular the similarity of their magnitudes  even at different times during the simulation when $\langle \partial(e_{\|}f_{i})/\partial t\rangle$ differs\,---\,suggests 
that parallel-electric-field work, \emph{viz.,} Landau damping, is responsible for the formation of the beam.


\begin{singlespace}

\section*{Acknowledgments}
We thank Bill Dorland, Ben Chandran, and Alfred Mallet for illuminating discussions. J.S. and R.M acknowledge support from the Royal Society Te Ap\=arangi, New Zealand through Marsden Fund grant UOO1727 and Rutherford Discovery Fellowship RDF-U001804. M.W.K. and E.Q. were supported by  the Department of Energy through the NSF/DOE Partnership in Basic Plasma Science and Engineering, awards DE-SC0019046 and DE-SC0019047, with additional support for E.Q. from a Simons Investigator Award from the Simons Foundation. L.A. acknowledges the support of the Institute for Advanced Study, and the work of A.A.S. was supported in part by UK EPSRC grant EP/R034737/1.
This research was part of the Frontera computing project at the Texas Advanced Computing Center, which is made possible by National Science Foundation award OAC-1818253. Further computational support was provided by the New Zealand eScience Infrastructure (NeSI) high performance computing facilities, funded jointly by NeSI's collaborator institutions and through the NZ MBIE, and through PICSciE-OIT TIGRESS High Performance Computing Center and Visualization Laboratory at Princeton University. The funders had no role in study design, data collection and analysis, decision to publish or preparation of the manuscript.

%

\end{singlespace}


%
%


\end{document}